\documentclass[11pt,a4paper]{article}
\usepackage{jheppub}

\def\beq{\begin{equation}}
\def\eeq{\end{equation}}
\def\bea{\begin{eqnarray}}
\def\eea{\end{eqnarray}} 

\newcommand{\newc}{\newcommand}
\newc{\Wp}{W^+}
\newc{\Wm}{W^-}
\newc{\Wpm}{W^{\pm}}
\newc{\ttbar}{t\bar{t}}

\usepackage{subfigure}
\newcommand{\zvv}{$Z\to\nu\bar{\nu}$~}
\newcommand{\pyt}{{\sc{Pythia8}}}
\newcommand{\gamb}{{\sc{Gambos}}}

\newcommand{\eg}{e.g.~}
\newcommand{\ie}{i.e.~}

\title{\boldmath Using $\gamma + $jets production to calibrate the Standard Model $ Z (\to \nu \bar\nu) +$jets background to new physics processes at the LHC}
\author{S. Ask,}
\author{M. A. Parker,}
\author{T. Sandoval,}
\author{M. E. Shea}
\author{and W. J. Stirling} 
\affiliation{Cavendish Laboratory, University of Cambridge, CB3 0HE, UK.}

\abstract{The irreducible background from $Z(\rightarrow \nu \nu)$+jets, to beyond the 
Standard Model searches at the LHC, can be calibrated using $\gamma$+jets data. 
The method utilises the fact that at high vector boson $p_T$ ($\gg M_Z$), the 
event kinematics are the same for the two processes and the cross sections differ 
mainly due to the boson--quark couplings. The method relies on a precise prediction 
from theory of the $Z/\gamma$ cross section ratio at high $p_T$, which should be 
insensitive to effects from full event simulation. We study the  $Z/\gamma$ ratio 
for final states involving 1, 2 and 3 hadronic jets, using both the leading--order 
parton shower Monte Carlo program \pyt\ and a leading--order matrix element program 
\gamb. This enables us both to understand the underlying parton dynamics in both 
processes, and to quantify the theoretical systematic uncertainties in the ratio 
predictions. Using a typical set of experimental cuts, we estimate the net theoretical 
uncertainty in the ratio to be of order $\pm 7\%$, when obtained from a Monte-Carlo 
program using multiparton matrix--elements for the hard process. Uncertainties associated with 
full event simulation are found to be small. The results indicate that an overall 
accuracy of the method, excluding statistical errors, of order $10\%$ should be 
possible.}
  
\arxivnumber{1107.2803}

\begin{document}

\maketitle

\section{Introduction}

At the CERN Large Hadron Collider (LHC), $Z(\to \nu\bar\nu)+$jets is an 
important Standard Model (SM) background to new physics processes that 
give rise to missing transverse energy + jets signals. In principle, the 
related process $Z(\to e^+e^-,\mu^+\mu^-)+$jets provides a way to calibrate 
this background, although in practice the number of such events may 
be too small to do this with sufficient precision. It has therefore been 
proposed \cite{atlas:susya,atlas:susyb,cms:susya,cms:susyb,cms:susyc} 
to use a related calibration process, $\gamma +$jets production, which has 
a much higher rate.\footnote{The same technique has also been used by the 
CDF collaboration \cite{cdf:vv} to estimate a similar background in weak 
boson pair production in a hadronic final state at the Tevatron.} The key 
point is that at high transverse momentum, $p_T \gg M_Z$, $\gamma$ and $Z$ 
production are very similar; indeed the only expected difference in rate 
comes from the different electroweak couplings of photons and $Z$ bosons 
to quarks, which are of course very well determined. There is also  no 
branching ratio suppression for photon production. The {\em measured} rate 
of  $\gamma + $jets production, coupled with theoretical knowledge of the 
{\em ratio} of $Z$ and $\gamma$ SM cross sections, can therefore be used 
to accurately predict the $Z(\to \nu\bar\nu)+$jets background. 

Although the $Z$ and $\gamma$ cross sections have a simple theoretical 
relationship at high vector boson $p_T$, care is needed when estimating the 
theoretical ratio of $Z$ and $\gamma$ SM cross sections. Especially for more 
than one jet, there are matrix element contributions to the cross section 
that may not be included in parton shower Monte Carlos such as {\sc{Pythia}} 
\cite{Sjo:06,Sjo:08} or {\sc{Herwig}} \cite{Bar:08}. In addition, it is important to 
quantify the theoretical {\it uncertainty} on the ratio (from the choice of 
PDFs, QCD scales etc.), since this will propagate through to the overall 
uncertainty on the background estimate. A detailed theoretical study of the 
$V + $~jets ($V=\gamma, Z$) cross sections and uncertainties is therefore 
required, to supplement the information from Monte Carlo event simulation. 
Finally, it is important to establish how well the theoretical precision 
survives under conditions closer to the experimental analysis, \eg including 
effects from full event simulation, jet reconstruction, detector acceptance, 
experimental cuts etc.

In this paper we report on such a study. Since our main interest is in estimating 
the missing $E_T$ distribution in events with multijets, we focus primarily 
on the inclusive vector--boson ($Z$ or $\gamma$) transverse momentum ($p_T$) 
distributions, particularly at high $p_T \gg M_Z$. We first analyse the ratio of 
the $Z+1$~jet and $\gamma +1$~jet $p_T$ distributions from a general theoretical 
perspective using a program for up to and including 3 jets based on exact leading--order 
parton--level matrix elements. We then reproduce these results using \pyt\ \cite{Sjo:06,Sjo:08} 
(at parton level). This establishes that the ratio of the two process cross sections 
is theoretically robust, particularly at high $p_T$. We predict the value of the 
cross section ratio using `typical' experimental cuts, and estimate its theoretical 
uncertainty. We then consider the corresponding $2$-- and $3$--jet cross sections, 
again comparing the exact leading--order matrix element results with those obtained 
from \pyt. Finally, we use our results to assess the systematic uncertainties on 
the missing transverse energy + jets background obtained from the photon + jets cross 
section using this method. 
The ratios predicted by the two alternative approaches, based on ``pure'' matrix 
elements or parton showers, are illustrated as well as used to constrain the systematic 
uncertainties related to the commonly used, and also more optimal, scenario where 
matrix elements are matched with the parton shower used in the MC simulation.
In the following sections, $V$ refers to the vector bosons 
$Z$ or $\gamma$.

In a recent paper \cite{Bern:2011pa}, a similar study was carried out for the $V+2$~jets 
cross sections using two approaches: next--to--leading order in pQCD applied at parton 
level, and exact leading--order matrix elements interfaced with parton showers (ME+PS) 
as implemented in {\sc{Sherpa}} \cite{Gle:09}. Agreement between our results and those 
of \cite{Bern:2011pa} confirms that over much of the relevant phase space, and particularly 
for inclusive quantities,  the effect of higher--order corrections on the leading--order 
$Z, \gamma$ cross section ratios is small. In our study, we consider also the $1$-- and 
$3$--jet ratios  and, more importantly, we extend the analysis to hadron level using \pyt. 
This allows for an analysis similar to the ones within the LHC experiments. We also 
examine the dependence of the ratios on the parton distribution functions, since these 
receive different weightings in the $Z$ and $\gamma$ cross sections and therefore do not 
exactly cancel in the ratio. Of course ultimately one would wish to evaluate all these 
cross sections consistently at next--to--leading order, using the methods described in 
\cite{Bern:2011pa} for the $2$--jet case.

\section{$V + $jets production in leading--order perturbative QCD} 
\label{sec:gambos}
 
\begin{figure}[ht]
\begin{center}
\vskip 0.5cm
\includegraphics[width=0.55\textwidth]{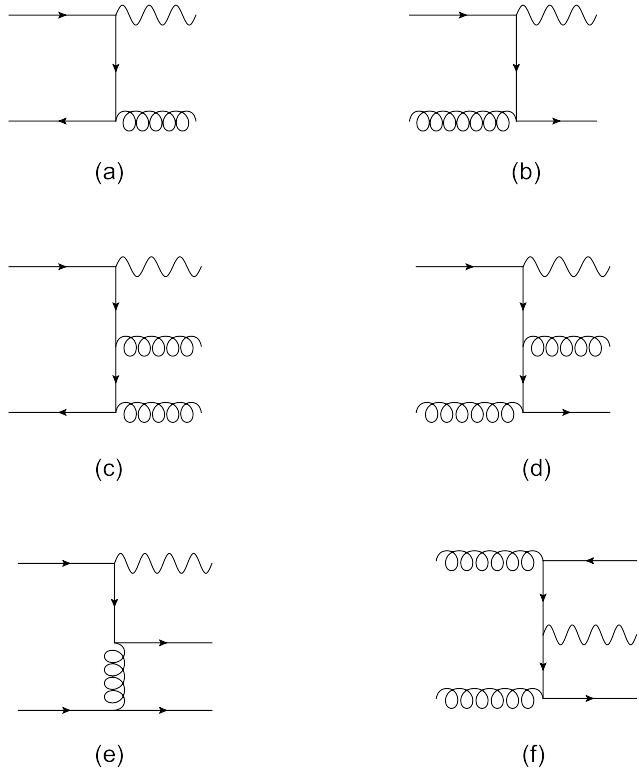}
\caption{Sample Feynman diagrams for $V + 1,2$~jets production, where $V=\gamma, Z$.}\label{fig:diags}
\end{center}
\end{figure}

In the Standard Model, the coupling of photons and $Z$ bosons to quarks $q$ 
are, respectively,
\beq
-ieQ_q \gamma^\mu\qquad \mbox{and} \qquad \frac{-ie}{2 \sin\theta_W \cos\theta_W} \gamma^\mu (v_q - a_q \gamma_5),
\eeq
where $Q_q$, $v_q$ and $a_q$, are respectively the electric, vector and axial 
neutral weak couplings of the quarks, and $\theta_W$ is the weak mixing angle. 
For hadron collider processes such as $q \bar q \to V + ng$ or $q g \to q V +(n-1)g$, 
both of which contribute to $V+n$~jets production, the matrix elements squared 
will contain factors of $Q_q^2$ or $(v_q^2 +a_q^2)/4\sin^2\theta_W\cos^2\theta_W$ 
for $\gamma$ or $Z$ respectively. The only other difference in the matrix 
elements comes from the non--zero $Z$ mass\footnote{For the purposes of this 
discussion, we treat the $Z$ as an on--shell stable particle. In practice, $Z$ 
decay will also form part of the matrix elements.}, which will appear in the 
internal propagators and phase space integration. Sample Feynman diagrams for 
$V+1$~jet production are shown in figures~\ref{fig:diags}(a,b), and for $V+2$~jet 
production in figures~\ref{fig:diags}(c--f).

\begin{figure}[ht]
\begin{center}
\subfigure[][]{\label{fig:ud:ud} \includegraphics[width=0.48\textwidth]{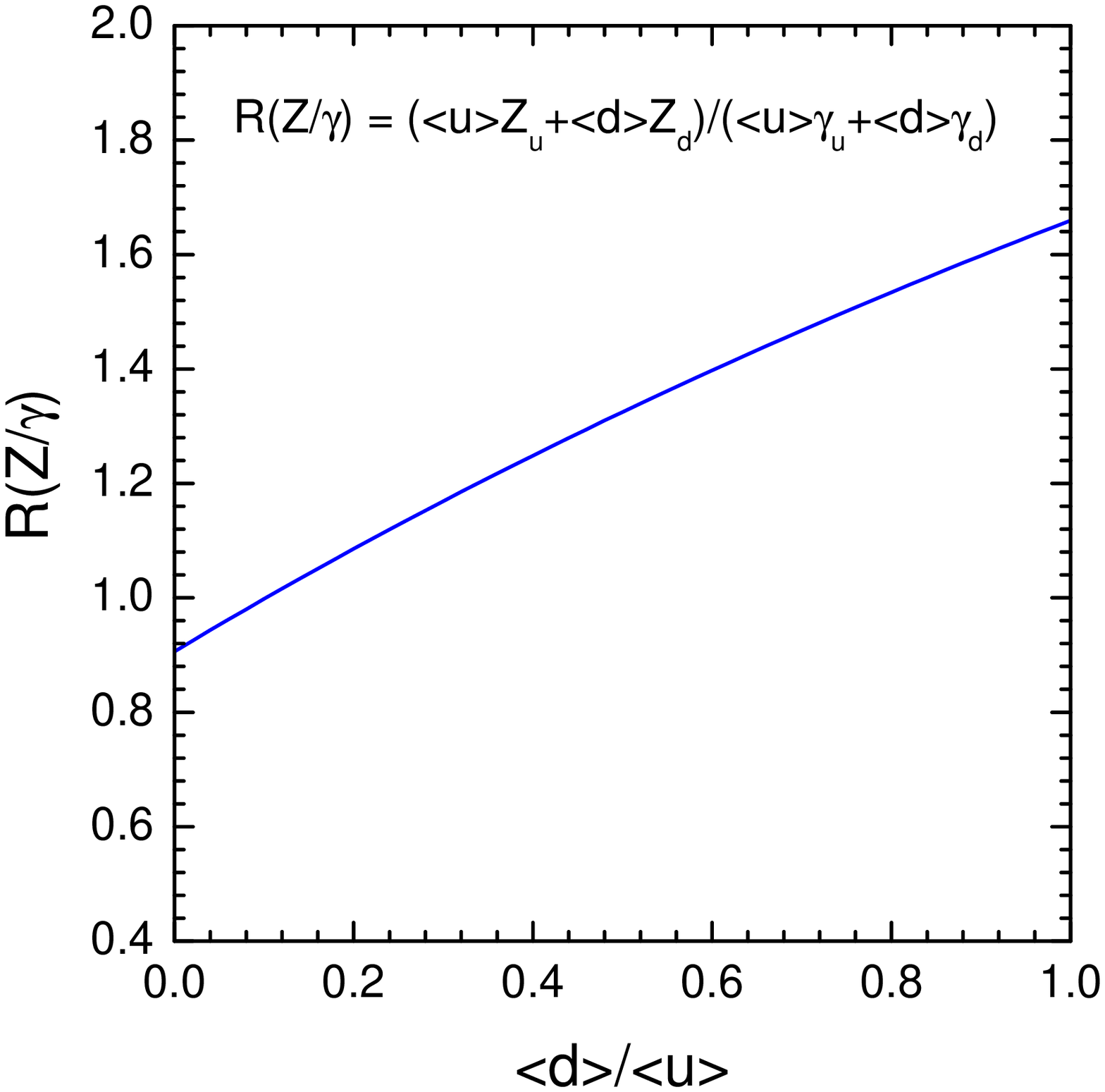}}
\subfigure[][]{\label{fig:ud:loud} \includegraphics[width=0.48\textwidth]{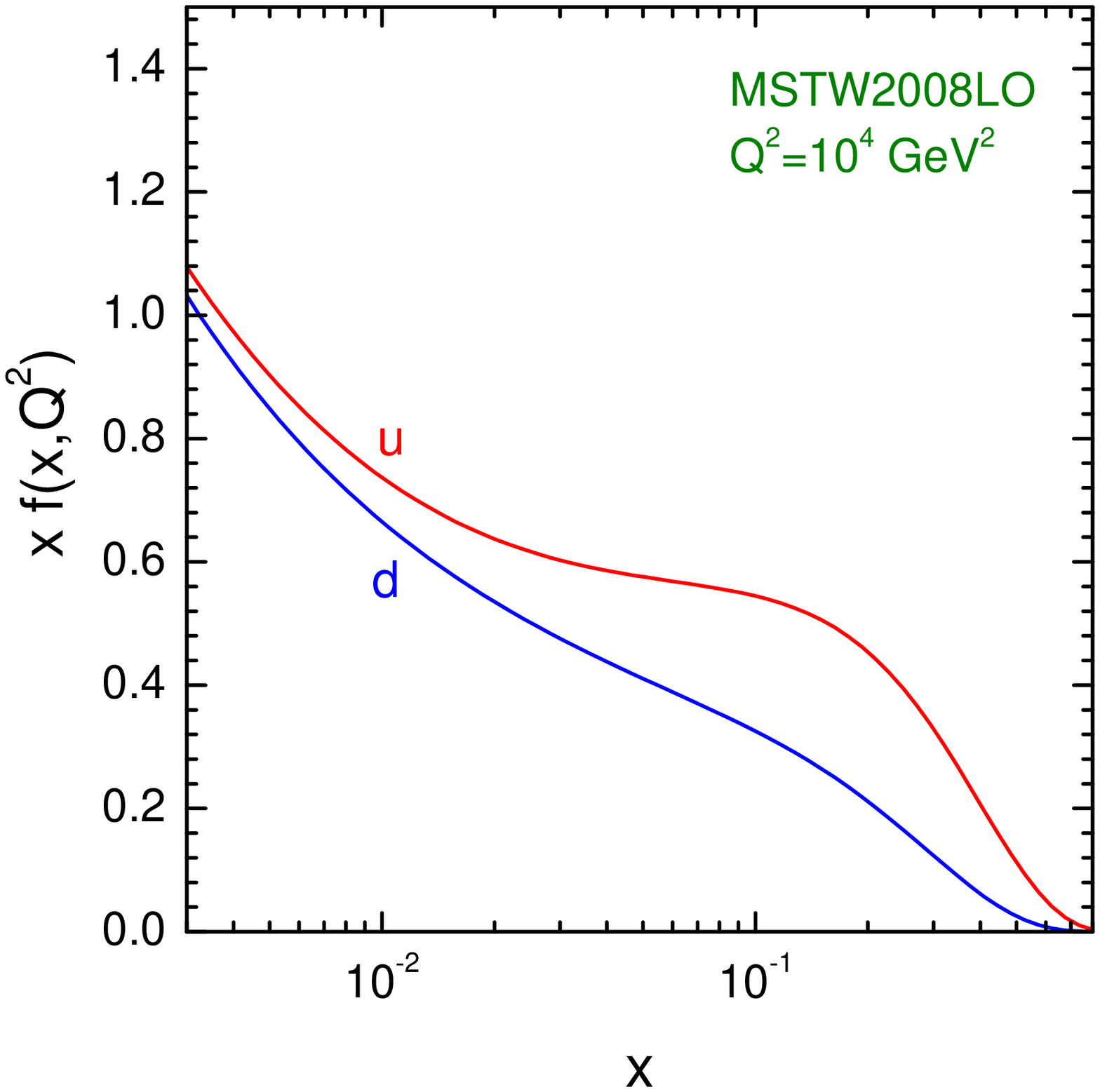}}
\caption{\subref{fig:ud:ud} Dependence of $R = \sigma(Z)/\sigma(\gamma)$ at the 
coupling constant level on the ratio of average $d$ and $u$ parton distribution 
values, see eq.~\ref{eq:ud}. \subref{fig:ud:loud} $u$ and $d$ quark MSTW2008LO PDFS 
\cite{Martin:2009iq} at $Q^2 = 10^4$~GeV$^2$.}\label{fig:ud}
\end{center}
\end{figure}

For high $p_T(V)$ ($\gg M_Z$) production we would therefore expect the $Z$ 
and $\gamma$ cross sections to be in the ratio
\beq
R_q = \frac{v_q^2 + a_q^2}{4 \sin^2\theta_W \cos^2\theta_W Q_q^2}. 
\eeq
Substituting $\sin^2\theta_W = 0.2315$, we obtain $R_u = 0.906$ and $R_d = 4.673$. 
In practice, of course, the cross sections will receive contributions from 
{\it all} quark flavour types, and so $R = \sigma(Z)/\sigma(\gamma)$ will be 
a weighted average of the $R_u$ and $R_d$ values, \ie 
\beq
R = \frac{Z_u \langle u \rangle + Z_d \langle d \rangle}{\gamma_u \langle u \rangle + \gamma_d \langle d \rangle}
\label{eq:ud}
\eeq
in an obvious notation, where $\langle u \rangle$ and $\langle d\rangle$ are 
the typical values of the $u$--type and $d$--type quark parton distribution 
functions (PDFs) in the cross section. Figure~\ref{fig:ud:ud} shows $R$ as a function 
of the ratio $\langle d \rangle / \langle u\rangle$. We would expect that where 
large $x$ values are probed, for example at very high $p_T(V)$, the ratio would 
approach the $R_u$ value since $d(x)/u(x) \to 0$ as $x \to 1$, see figure~\ref{fig:ud:loud}. 
For moderate $p_T$ values at the LHC, $\langle x \rangle \sim 0.1$, which 
corresponds to $\langle d \rangle / \langle u\rangle \simeq 0.6$ and therefore 
$R \simeq 1.4$.

The simple connection (\ref{eq:ud}) between the vector boson cross section
ratio and the initial state quark flavour is, however, broken for $n_{\rm
jets} \geq 2$. Consider for example the sample Feynman diagrams of 
figure~\ref{fig:diags}(e) and (f). For the former `four--quark' diagrams, the
vector boson can be emitted off any of the external quark legs and so the
numerator and denominator of the ratio $R$ depend on more complicated
products of quark distributions. Because at high $x$ $uu$ scattering will 
be relatively more dominant than $dd$ scattering, we would expect that the 
value of $R$ for such processes would be closer to $R_u$ than to $R_d$, 
compared to the $1$--jet ratio. On the other hand, for the $gg$--scattering 
diagrams, figure~\ref{fig:diags}(f), the ratio of the corresponding cross 
sections is (ignoring the $Z$ mass) $R = \sum_q Z_q / \sum_q \gamma_q$, 
where the sum is over the final state quark (antiquark) flavours, and the 
dependence on the initial state (gluon) distributions cancels. By way of 
illustration, with 5 massless flavours we obtain $R = 1.933$. As we shall 
see below, the four--quark contribution is more important at high $p_T$ than 
the $gg$ contribution, and the net effect is to reduce $R$ slightly compared 
to the $1$--jet case.

\begin{figure}[ht]
\begin{center}
\subfigure[][]{\label{fig:gam:pt} \includegraphics[width=0.48\textwidth]{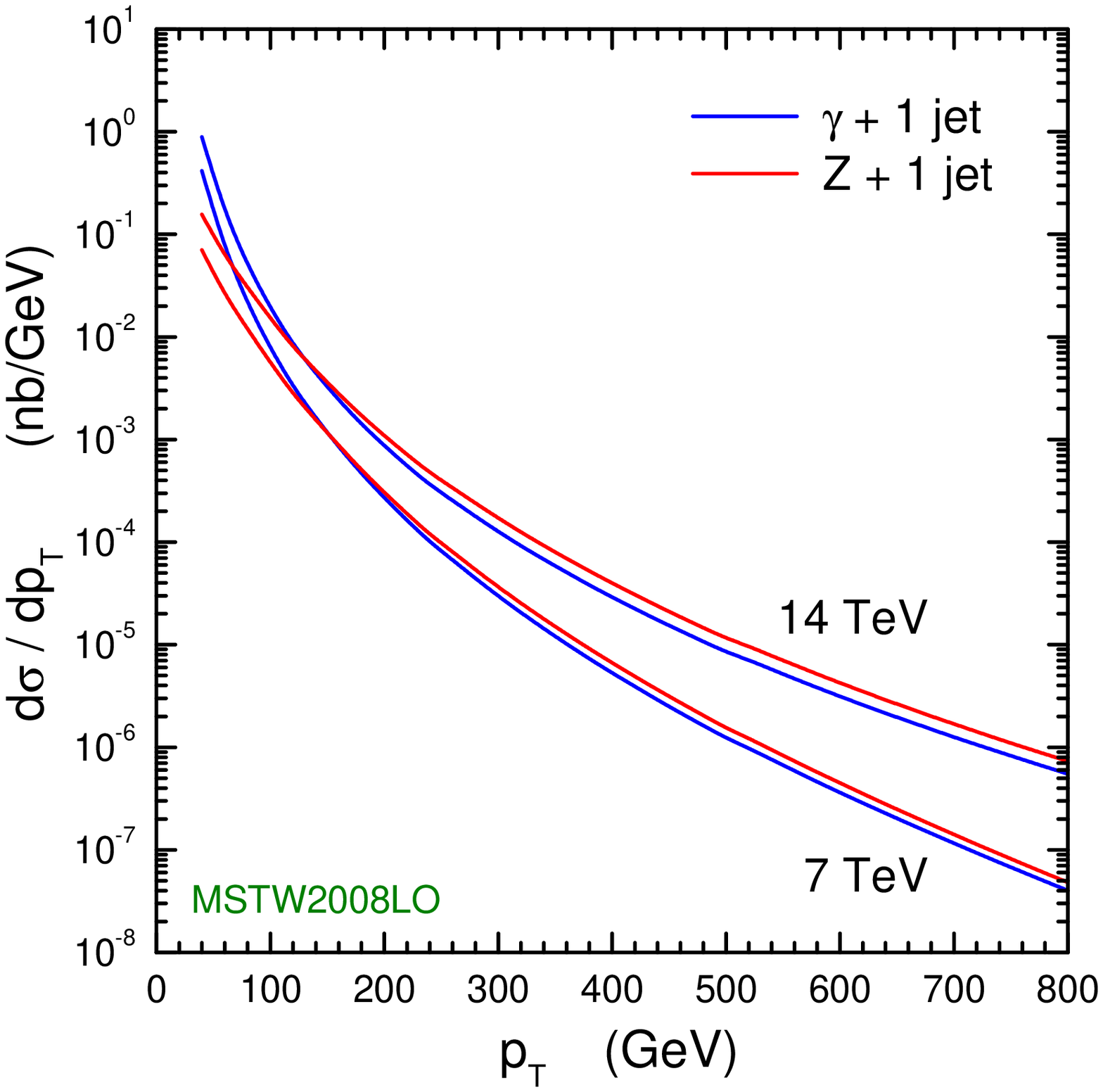}}
\subfigure[][]{\label{fig:gam:rat} \includegraphics[width=0.48\textwidth]{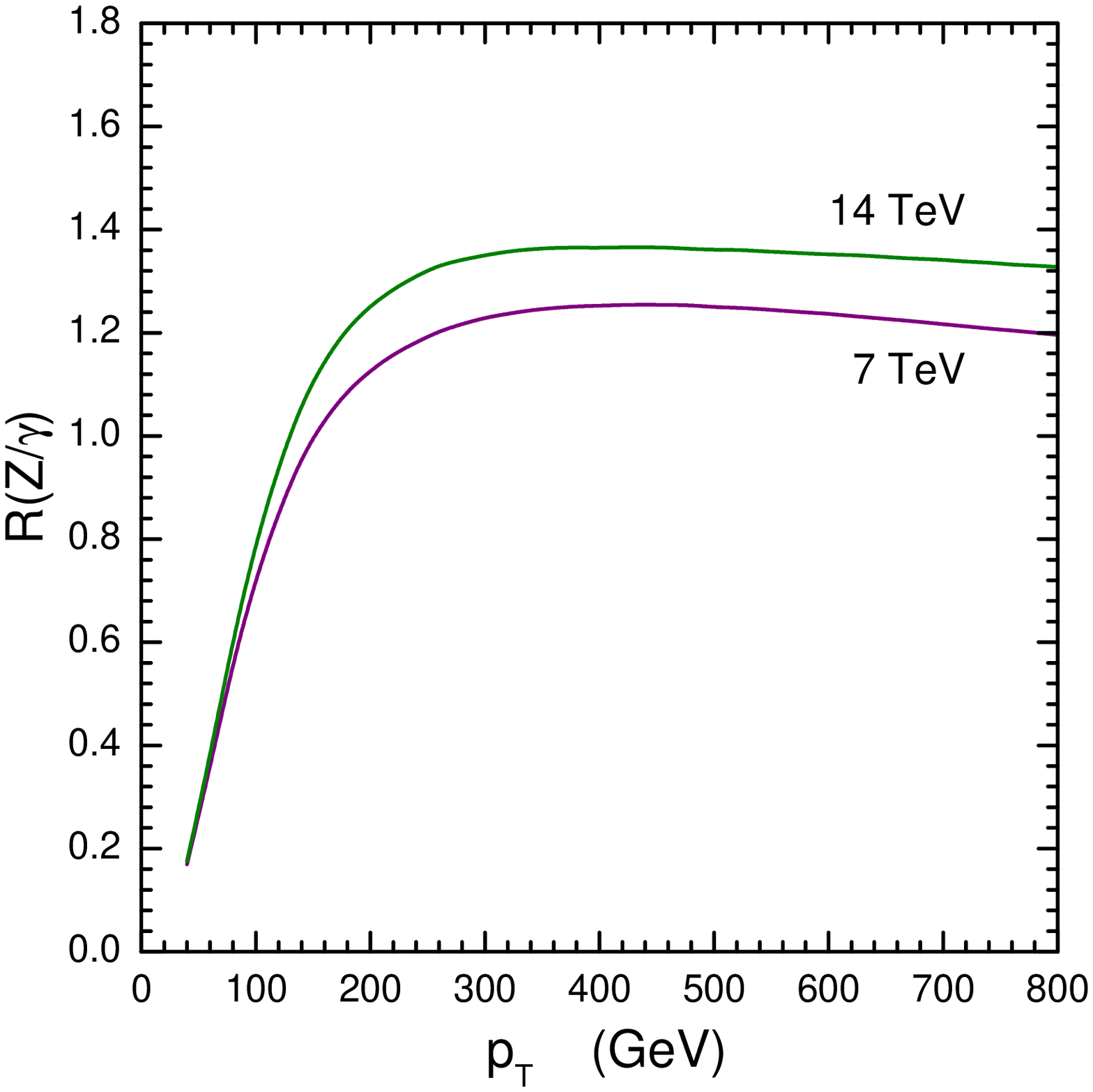}}
\caption{ \subref{fig:gam:pt} $Z$ and $\gamma$ $p_T$ distributions at $\sqrt{s}= 7$~TeV 
and 14~TeV, with parameters and cuts as described in the text. \subref{fig:gam:rat} Ratio 
of the $Z$ and $\gamma$ $p_T$ distributions.}\label{fig:gambosZgpt}
\end{center}
\end{figure}

Figure~\ref{fig:gam:pt} shows the $Z+1$~jet and $\gamma+1$~jet cross sections\footnote{Expressions 
for the matrix elements can be found, for example in Chapter~9 of \cite{QCDbook}.} 
as functions of the vector boson transverse momentum at $\sqrt{s}= 7$~ TeV and 
14~TeV. Standard PDG values of the electroweak parameters are used, and the PDFs 
are the leading--order MSTW2008LO set \cite{Martin:2009iq} with renormalisation 
and factorisation scale choice $\mu_R = \mu_F = p_T(V)$, and the $Z$ is treated 
as an on--shell stable boson. The acceptance cuts are $\vert y(V,j)\vert < 2.5$  
and $p_T(V,j) > 40$~GeV, where $y$ is the rapidity. Figure~\ref{fig:gam:rat} shows 
the ratio of the $Z$ and $\gamma$  distributions. We see the expected behaviour 
of a roughly constant ratio at large $p_T(V) \gg M_Z$ lying between the $R_u$ and 
$R_d$ values defined above. Although above $p_T \sim M_Z$ the ratio does exhibit 
a plateau region, at very large $p_T$ we begin to see a slight decrease, as  the 
high--$x$ behaviour of the $d/u$ PDF ratio drives the ratio down towards the $R_u$ 
value. At 14~TeV, the empirical large--$p_T$ value of $R \simeq 1.4$ is consistent 
with $\langle d \rangle / \langle u\rangle \simeq 0.6$, see figure~\ref{fig:ud:ud}. 
This in turn is consistent with $u$ and $d$ PDFs probed in the $x \sim 0.1$ region, 
see figure~\ref{fig:ud:loud}.  At the lower collider energy (7~TeV), higher $x$ values 
are sampled for the same $p_T$, and the $Z/\gamma$ ratio decreases slightly, moving 
towards the $R_u$ value. Note that the ratio curves in figure~\ref{fig:gam:rat} can be 
reasonably well approximated by
\beq
R = R_{0} \left(\frac{p_T^2}{p_T^2+M_Z^2}\right)^{n},
\label{eq:1Jratiofit}
\eeq
with $n \approx 1.2$, illustrating the expected $Z$ mass suppression relative to the 
photon distribution for $p_T < M_Z$. 

\section{Parton level analysis}
\label{sec:partonlevel}

In order to study the ratio of the $Z$ and $\gamma$ distributions in a realistic 
experimental environment, we need to use an event simulation Monte Carlo program. 
We use \pyt\ for this purpose. At the same time, we want to understand the difference 
between the scattering amplitudes embedded in \pyt\ and the amplitudes obtained 
using exact QCD matrix elements for multijet production. For the latter we use 
the program \gamb, an adaptation of the Giele et al. {\sc Vecbos} program 
\cite{VECBOS} for $W,Z + n$~jets production, in which the weak boson is replaced 
by a photon.

First, we compare the \pyt\ and \gamb\ results at the parton level. This serves to 
check the consistency of the results from the two programs, when configured as 
similarly as possible, and provides a common middle step between the matrix--element 
(ME) $V+$~jets results at parton level produced by \gamb\ and the results for fully 
simulated events from \pyt.

The \pyt\ results are obtained using the LO ($2 \rightarrow 2$) processes, 
$q \bar{q} \rightarrow V g$ and $ q g \rightarrow V q$ where $V = \gamma$ or $Z$, 
corresponding to the Feynman diagram types shown in figures~\ref{fig:diags}(a,b). 
Events with $\geq 2$ jets are generated by parton showering off the initial and 
final state partons. This means that processes such as those shown in figures~\ref{fig:diags}(c) 
to~\ref{fig:diags}(f) are included, albeit with an approximation to the exact matrix 
elements, and we can therefore expect differences between \gamb\ and \pyt\ results for 
the $Z/\gamma$ ratios with $\geq 2$ jets.

In  order to produce results directly comparable with \gamb, the following settings 
are used as default in \pyt:
\begin{itemize}
\item{{\tt PDFs:} MSTW2008LO;}
\item{{\tt Strong:} $\alpha_S(M_Z^2) = 0.13939$, with one loop running;}
\item{{\tt EM:} $\alpha _{EM}(M_Z^2) = 1/127.918$, with one loop running;}
\item{{\tt Weak:} $\sin^2(\theta_W) = 0.2315$;}
\item{{\tt Scales:} Renormalisation and factorisation scales, $\mu_{R} = \mu_{F} = p_T(V)$;}
\item{Rapidity, transverse momentum and separation cuts on the final--state $Z$, $\gamma$ 
and jets as described in the previous section.}
\end{itemize}
In the following sub--sections we first compare results for the $V+1$~jet distributions, 
then discuss the theoretical uncertainties on the corresponding $Z/\gamma$ ratio, 
and finally compare the $V+2,3$~jet results from the two programs.

\subsection{$V+1$~jet results} \label{sec:pyt1}

\begin{figure}[ht]
\begin{center}
\subfigure[][]{\label{fig:pyt:pteCM} \includegraphics[width=0.48\textwidth]{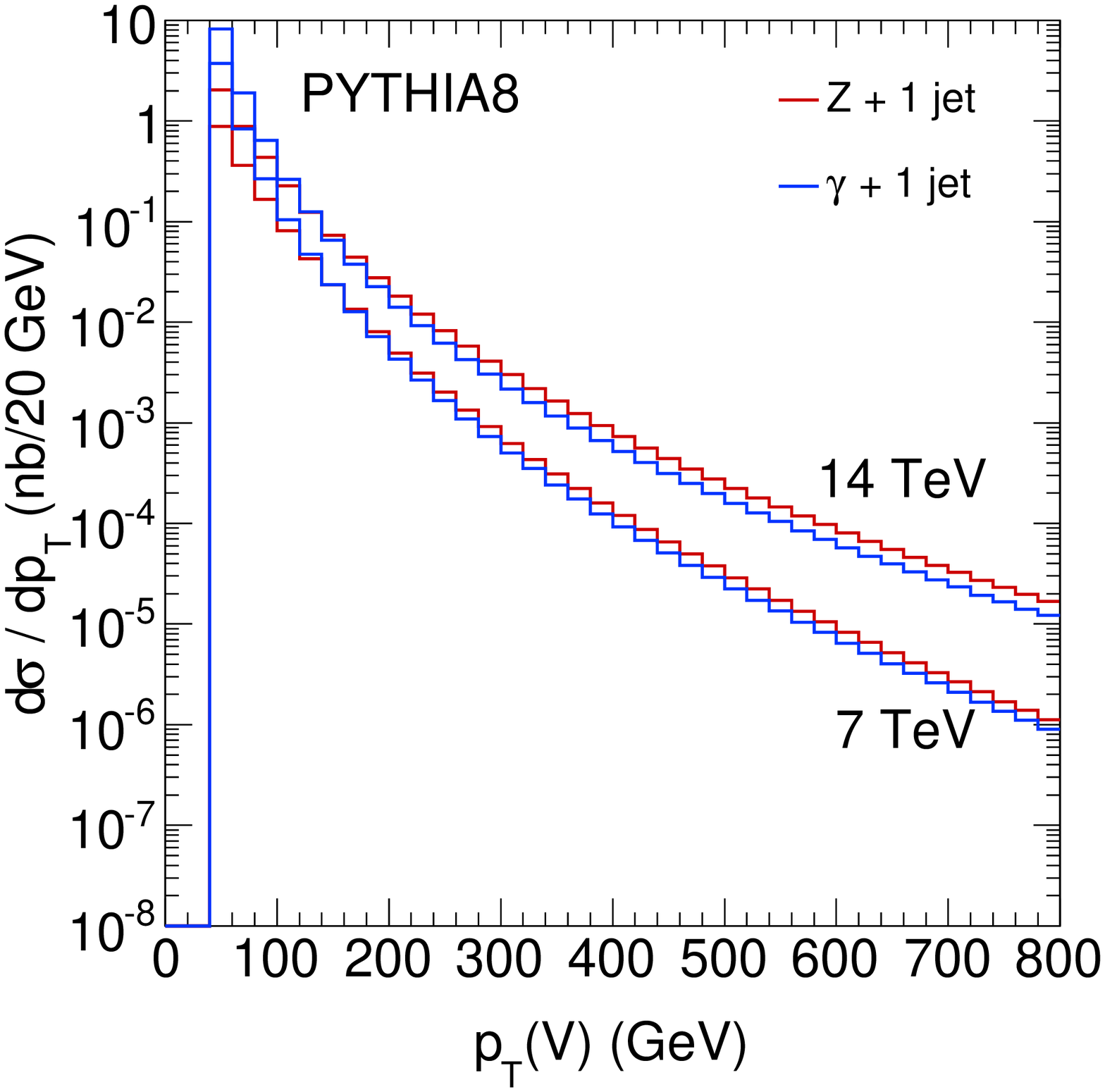}}
\subfigure[][]{\label{fig:pyt:rateCM} \includegraphics[width=0.48\textwidth]{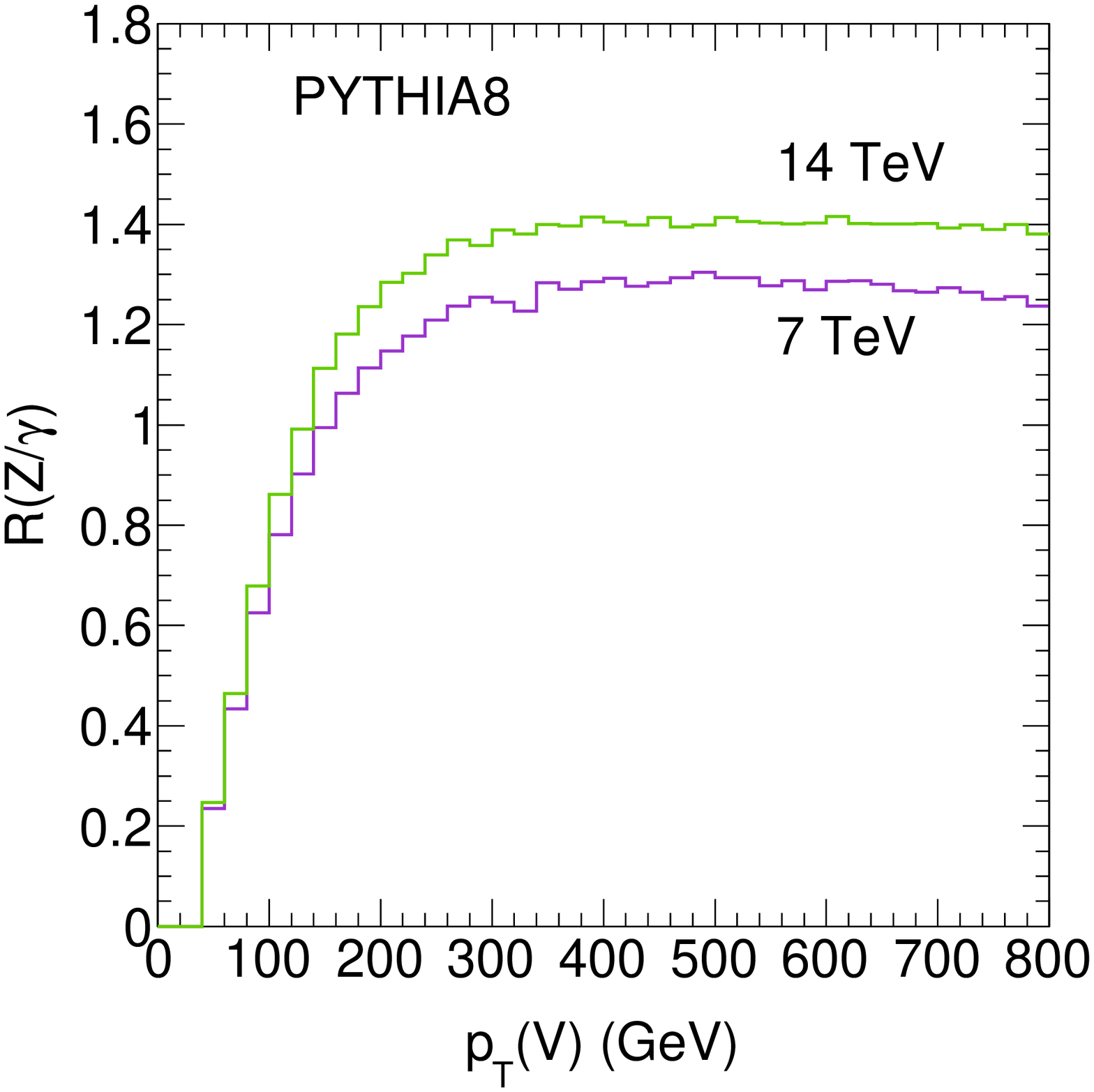}}
\caption{ 
\subref{fig:pyt:pteCM} Differential cross section as a function of the vector boson $p_T$ 
for the process $pp\to V+1$~parton with $V=\gamma, Z$ from \pyt, and \subref{fig:pyt:rateCM}
the ratio of these.
\label{fig:pyt:eCM}}
\end{center}
\end{figure}

\begin{figure}[ht]
\begin{center}
\includegraphics[width=0.48\textwidth]{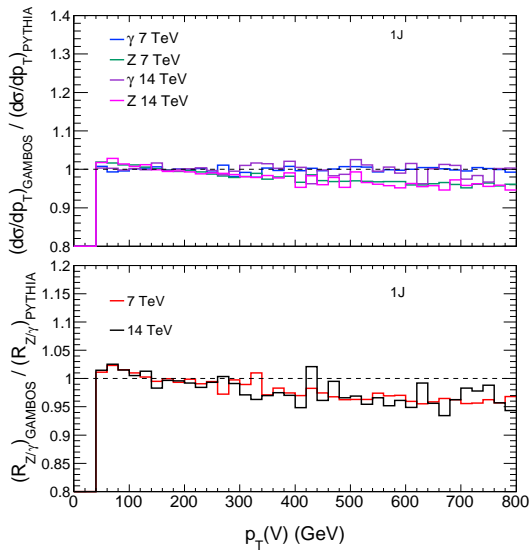}
\caption{Effect on the boson $p_T$ and the ratio from using a matrix--element generator 
(\gamb) and \pyt\ at 7 and 14 TeV. 
\label{fig:vareCM} }
\end{center}
\end{figure}

The LO matrix elements used in the \pyt\ processes are the same as those used in 
\gamb\ for the $V+1$~jet case, and a parton level comparison between the two programs 
should therefore give identical results. Note that  the $V$ and the jet correspond 
to the outgoing partons of the hard process in \pyt, without any further simulation 
of the event. 
The differential $Z$ and $\gamma$ cross sections and their ratio predicted by \pyt\ 
are shown in figure~\ref{fig:pyt:eCM} for $pp$ collisions at 7 as well as 14~TeV. The results show the 
same characteristic features already seen in the \gamb\ predictions in figure~\ref{fig:gambosZgpt}, 
\ie the $Z$ cross section, excluding any branching ratios, is smaller than the photon 
cross section at small $p_T$ due to the mass suppression, but is roughly proportional 
to, and slightly larger than, the $\gamma$ cross section  at $p_T \gg M_Z$.  The 
ratios obtained from \pyt\ and \gamb\ are compared in figure~\ref{fig:vareCM}, for 7~TeV 
and 14~TeV collision energies. Evidently there is good agreement between the two programs, 
as expected. As seen in the plot, the photon cross sections from the two programs 
agree perfectly, whereas a small difference between the $Z$ cross sections, $<$5\%, 
is visible. This difference is due to the way in which the \pyt\ generator treats the 
$Z$ boson as a resonance, in contrast to \gamb\ where the $Z$ is treated as a real 
particle, and this was confirmed by producing a \gamb\ like process in \pyt\ which 
reproduced the same results. Since the $Z$ boson is generally treated as a resonance 
in MC programs, this difference is not considered as a source of uncertainty.


\subsection{Theoretical uncertainties} \label{sec:theoryunc}

Having established numerical agreement between the two programs, we can use \pyt\ to 
investigate the theoretical uncertainty on the $Z/\gamma$ ratio at high $p_T$. There 
are a number of sources of these, which we address in turn.

\begin{figure}[ht]
\begin{center}
\subfigure[][]{\label{fig:pdfsloud} \includegraphics[width=0.48\textwidth]{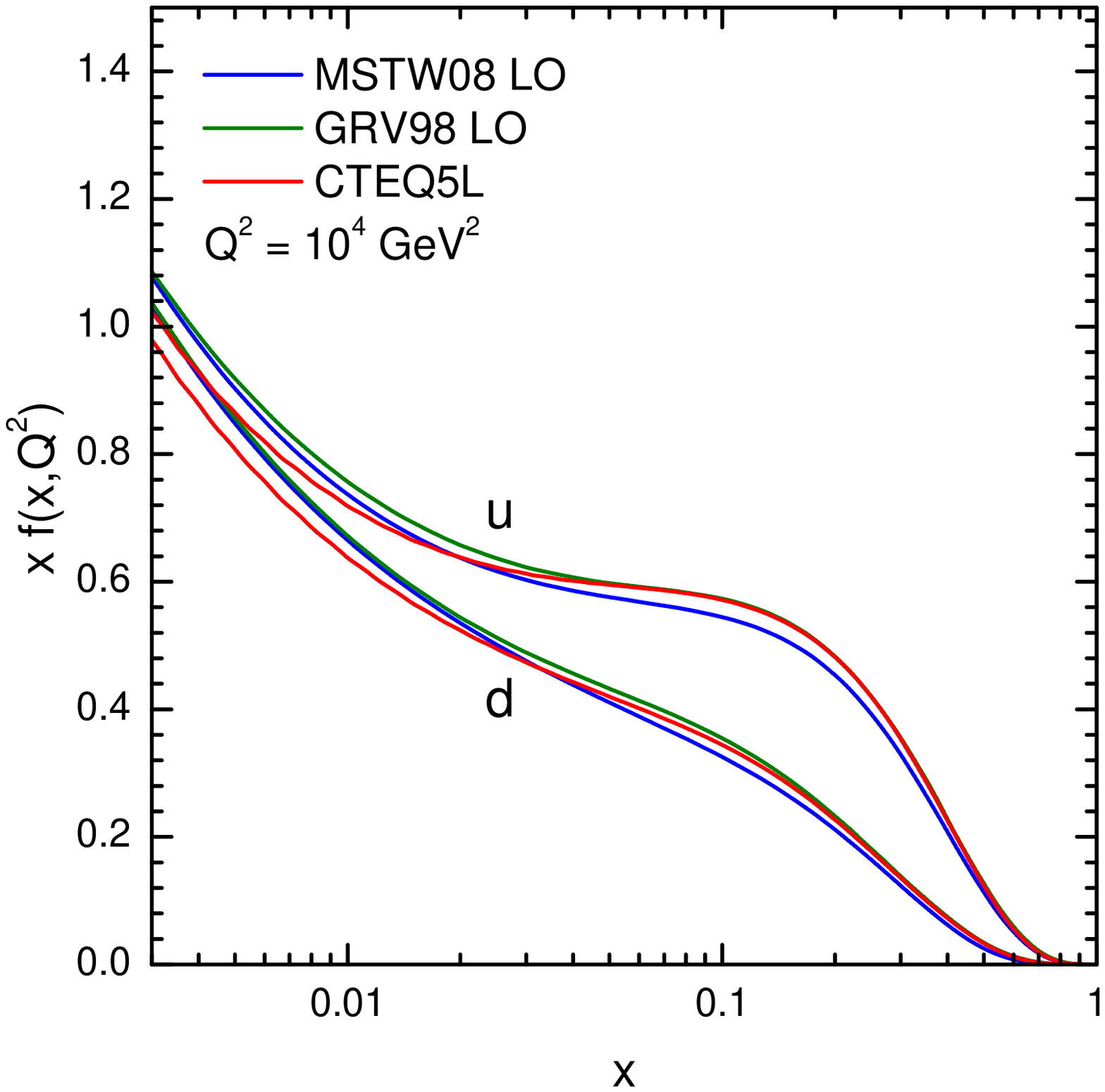}}
\subfigure[][]{\label{fig:pdfsloud2} \includegraphics[width=0.48\textwidth]{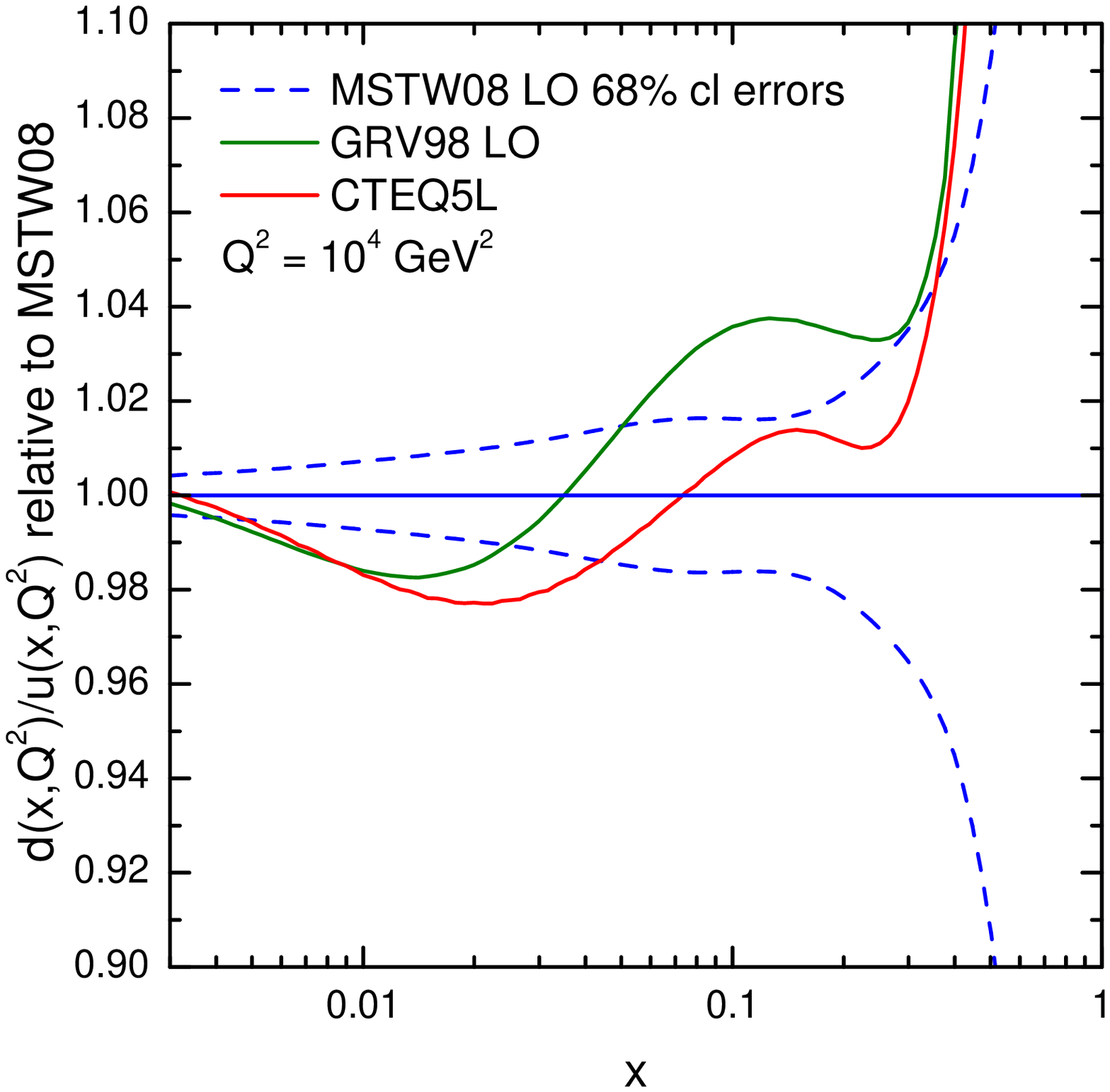}}
\caption{\subref{fig:pdfsloud} Comparison of up and down quark distributions from MSTW2008LO, 
CTEQ5L and GRV98. \subref{fig:pdfsloud2} $d/u$ ratios compared to MSTW2008LO. \label{fig:pdfsloudtot}}
\end{center}
\end{figure}

\begin{figure}[ht]
\begin{center}
 \includegraphics[width=0.48\textwidth]{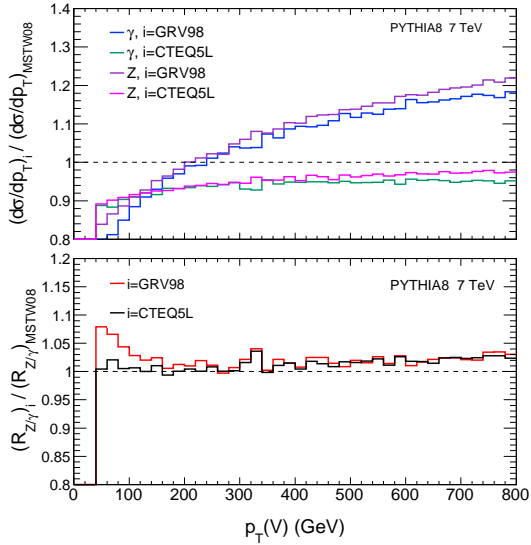}
\caption{Effects on ($Z,\gamma$) differential cross sections and cross section ratio 
after varying the PDFs}
\label{fig:varpdf}
\end{center}
\end{figure}

First, we consider the dependence on the PDFs used in the calculation. As argued in 
section~\ref{sec:gambos}, the $Z/\gamma$ ratio at high $p_T$ is sensitive to the $d/u$ 
parton ratio at large $x$. To study the possible variation in this ratio, we investigate 
the spread from using the different eigenvectors of the MSTW2008LO set and we compare 
these predictions with those from two (older) leading--order PDF sets, 
CTEQ5L \cite{Lai:00} and GRV98 \cite{Glu:98}, shown in figure \ref{fig:pdfsloudtot}. 
The latter should yield a {\it conservative} estimate of the PDF dependence. The impact on the 
$Z$ and $\gamma$ distributions and their ratio is shown in figure \ref{fig:varpdf}.
 Note that the CTEQ5L and GRV98 PDFs give respectively softer 
and harder $Z$ and $\gamma$ $p_T$ distributions, which is an artefact of the underlying 
quark and gluon PDF behaviour, see figure~\ref{fig:pdfsloud}, but that the effect largely cancels
in the ratio. The residual small differences in the cross section ratio can be understood in terms of 
the corresponding small differences in the $d/u$ ratio for the various sets, shown in 
figure~\ref{fig:pdfsloud2}. We ascribe a conservative $\pm 4$\% PDF uncertainty to  the  
$R(Z/\gamma)$ ratio at high $p_T(V)$. 

\begin{figure}[ht]
\begin{center}
\subfigure[][]{\label{fig:varrenormS} \includegraphics[width=0.48\textwidth]{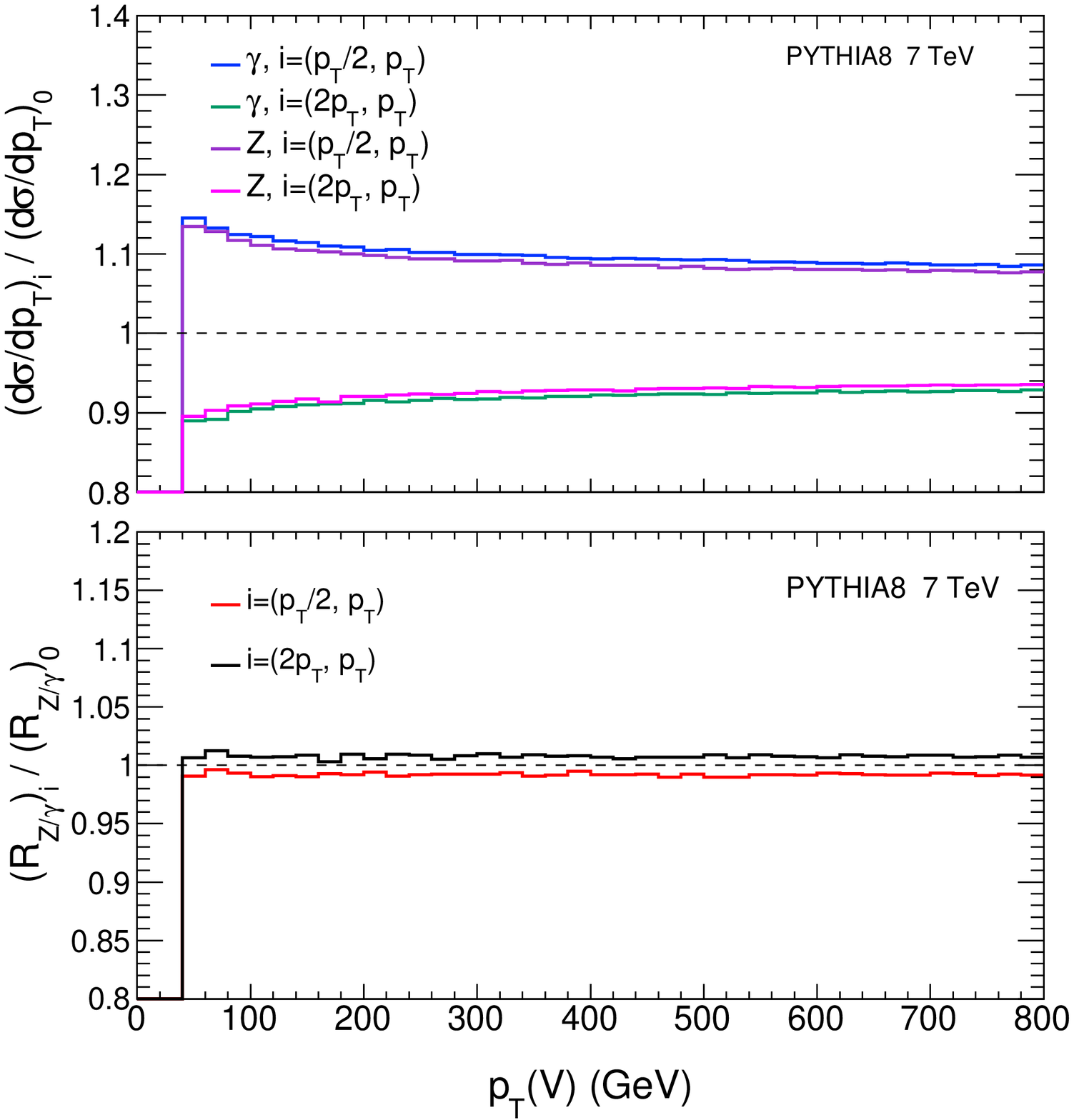}}
\subfigure[][]{\label{fig:varfactorS} \includegraphics[width=0.48\textwidth]{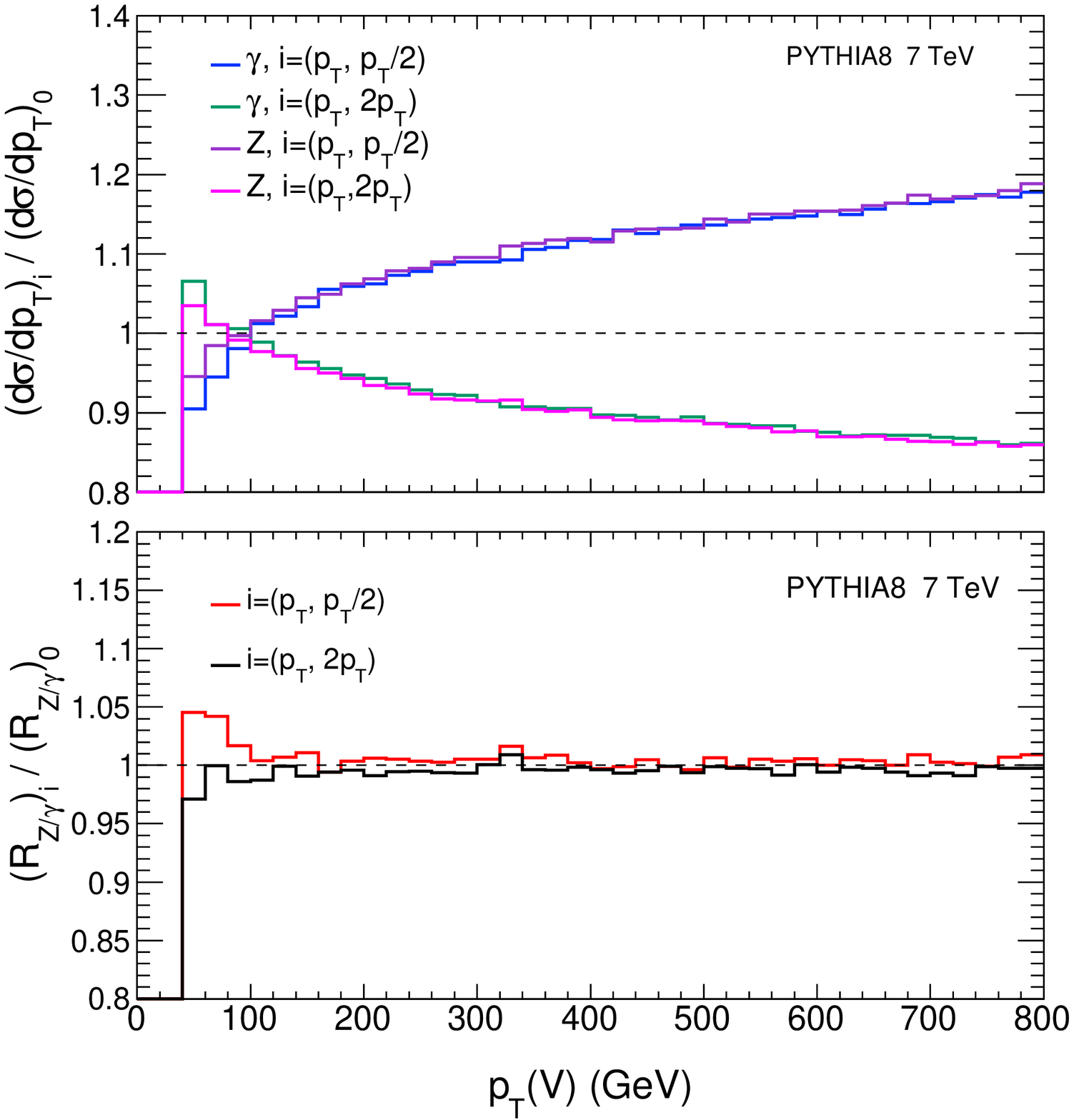}}
\subfigure[][]{\label{fig:varrenormC} \includegraphics[width=0.48\textwidth]{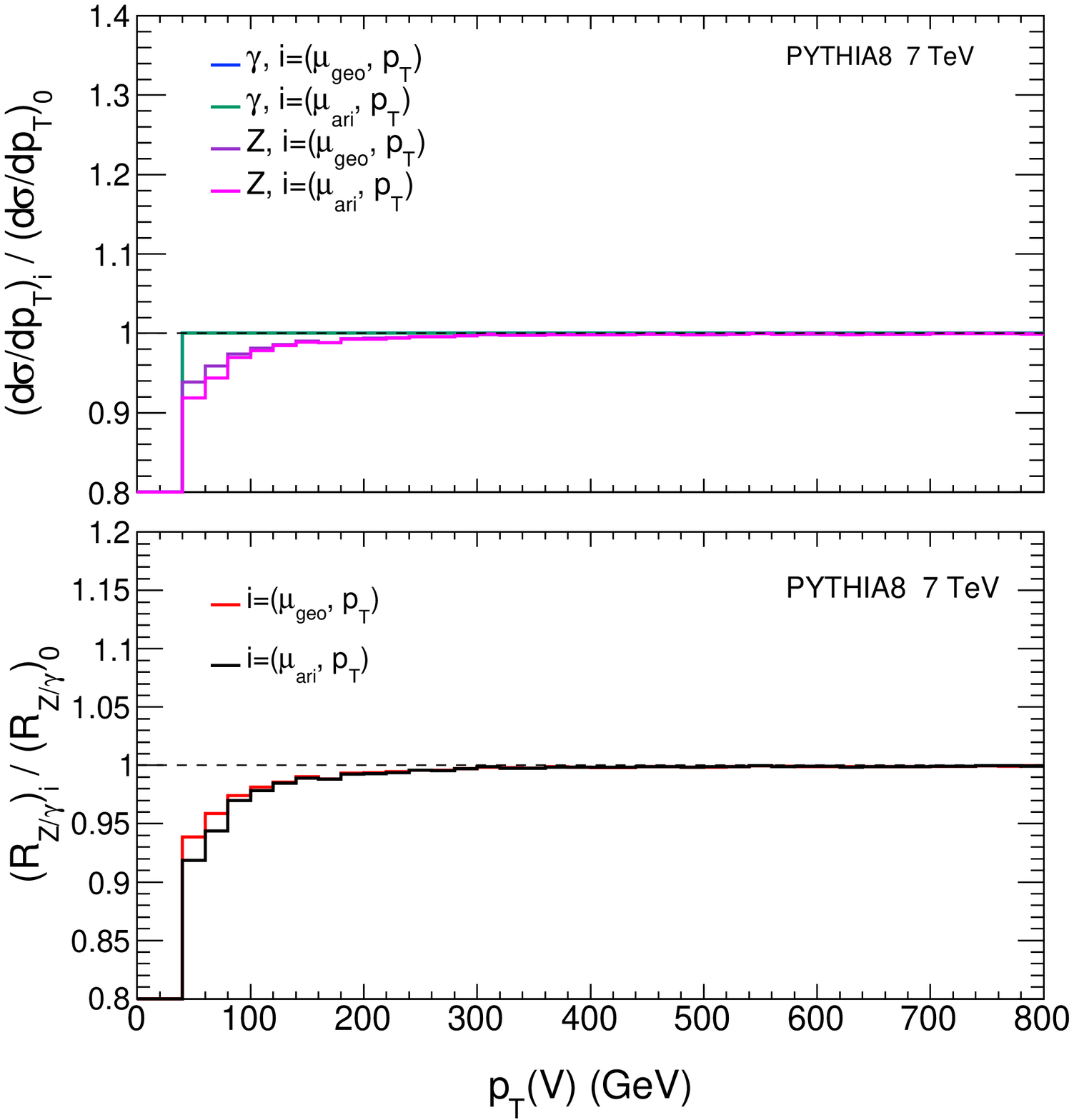}}
\subfigure[][]{\label{fig:varfactorC} \includegraphics[width=0.48\textwidth]{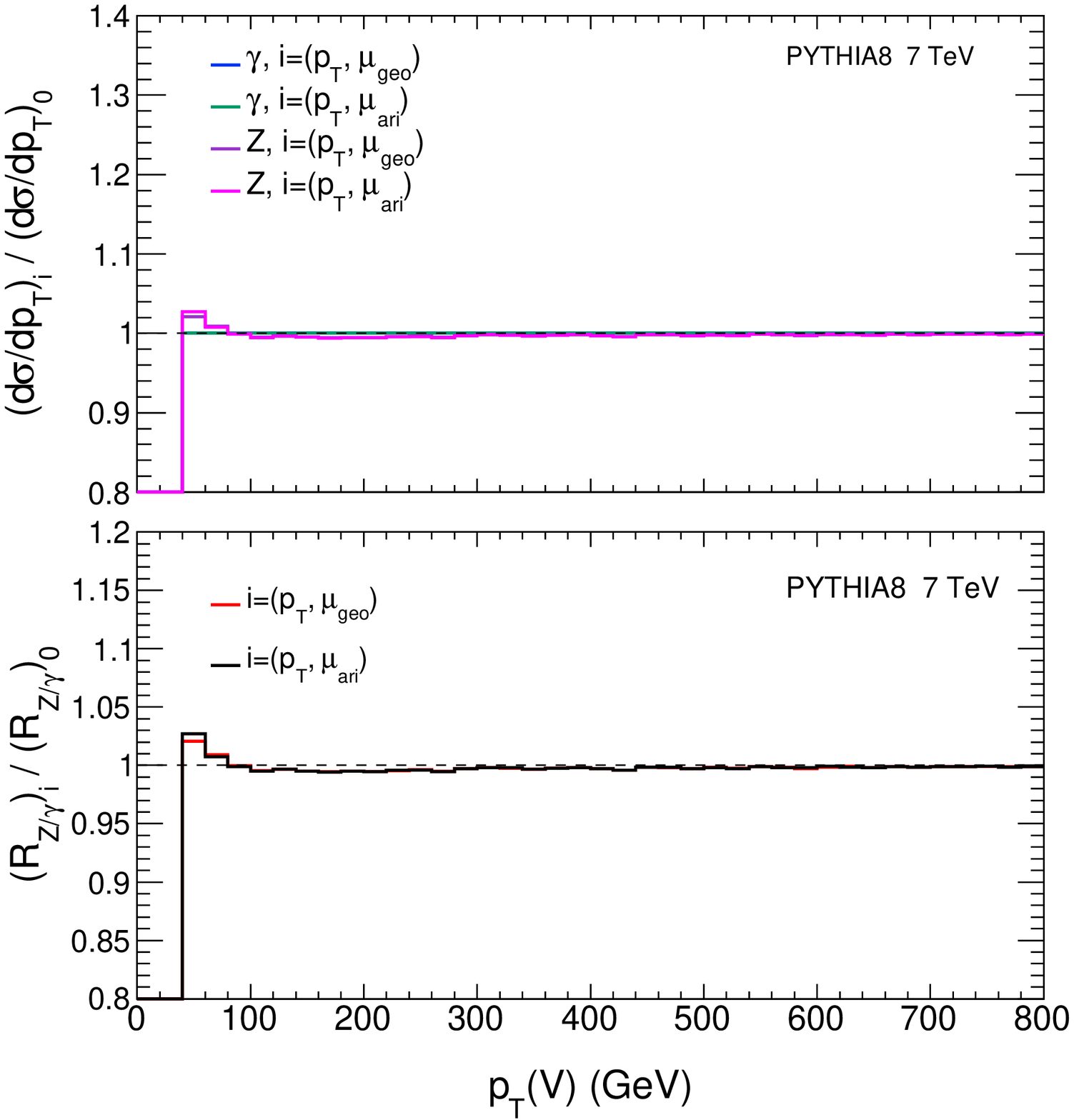}}
\caption{ Effects on ($Z,\gamma$) differential cross sections and cross section ratio 
after varying the scales ($\mu _R$, $\mu _F$). The different $\mu$ scales are defined 
in the text and the denominators correspond to using the default scale choice, i.e. $0 = (p_T,p_T)$. \label{fig:varAll}}
\end{center}
\end{figure}

In addition to varying the PDFs, we use different choices for the renormalisation 
and factorisation scales, $\mu_R$ and $\mu_F$, in order to mimic the effect of 
higher--order pQCD corrections not included in either the \pyt\ or \gamb\ analyses. 
In particular, we use multiples of the default scales $\mu_R = \mu_F = p_T(V)$, 
and two variants of this: the arithmetic and geometric means of the
final-state transverse masses in the $2\to 2$ hard process,
$\mu_{\rm ari}^2 = (m_{T1}^2 + m_{T2}^2)/2 = (2p_T(V)^2 + m(V)^2)/2$ and
$\mu_{\rm geo}^2 = m_{T1} m_{T2} = p_T(V) \sqrt{p_T(V)^2 + m(V)^2}$. Note that
for the photon, these scales are identical to the default scale $p_T(\gamma)$.
Figure~\ref{fig:varAll} presents the corresponding impact on the 
differential cross sections, $d\sigma/dp_T$, as well as the cross section ratio, 
$R(Z/\gamma)$. The results show that although the variations have significant effects 
on the differential cross sections, as expected, the $Z/\gamma$ ratio remains stable 
in the regime $p_T\gg M_Z$, and for $p_T(V) > 100$~GeV all variations of the ratio are 
within $\pm 3$\%.


Indeed the only sizable effect on the ratio related to the scales is observed from 
the different choices of the scale $\mu_{R}$, which becomes visible at $p_T \ll M_Z$. 
Any choice of scale of the form $\kappa p_T$ will of course cancel in the $Z/\gamma$ 
ratio, but scales of the form $\kappa \sqrt{p_T^2+M_V^2}$ will give different results 
for low $p_T \sim M_V$. The size of this effect was also shown to be consistent with 
the ratio $\alpha_S(\mu_R(Z))/\alpha_S(\mu_R(\gamma))$ using the same one--loop formula 
as in \pyt\ and \gamb. No similar effects are observed from different choices of the 
factorisation scale $\mu_{F}$, since the PDFs vary only weakly with the factorisation scale
at the $x$ values probed by these cross sections.

Note that in the above analysis we have used the {\it same} form of scale variation 
simultaneously in both the numerator ($Z$) and denominator ($\gamma$) cross sections. 
As pointed out in Ref.~\cite{Bern:2011pa}, this gives a much smaller scale variation 
than if the scales are varied {\it independently} in the two cross sections. However, 
we argue that if we select $Z$ and $\gamma$ events for which the kinematics of the 
(colour--singlet) vector bosons and the jets are the same, and if the energies and 
momenta are large enough such that the $Z$ mass can be neglected (\eg $p_T \gg M_Z$), 
then the higher--order pQCD corrections to both cross sections should essentially be 
the same and should therefore largely cancel in the ratio.

\subsection{$V+2,3$~jets results} \label{sec:pyt23}

\begin{figure}[ht]
\begin{center}
\subfigure[][]{\label{fig:gr:dr} \includegraphics[width=0.48\textwidth]{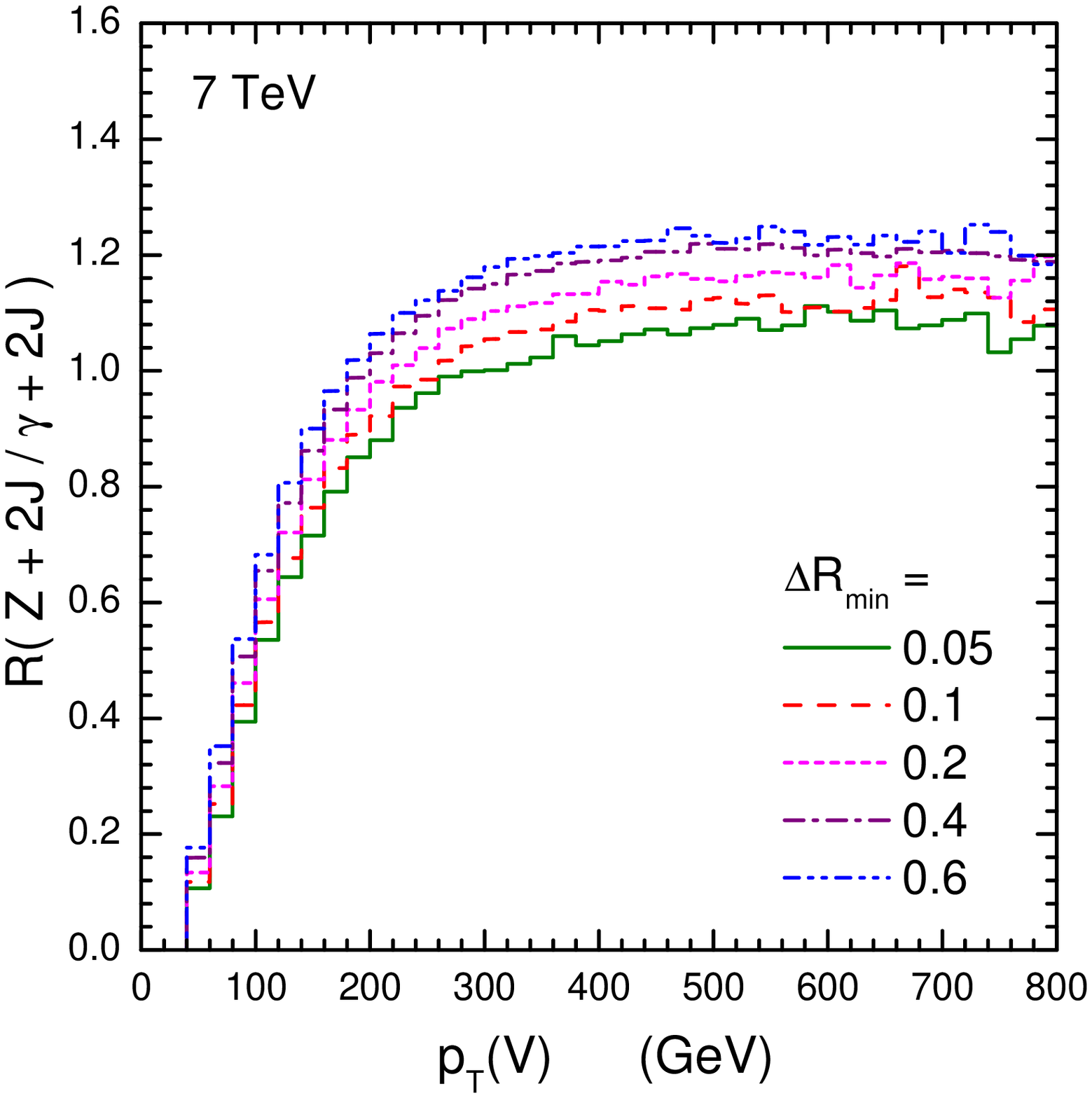}}
\subfigure[][]{\label{fig:gr:comp} \includegraphics[width=0.48\textwidth]{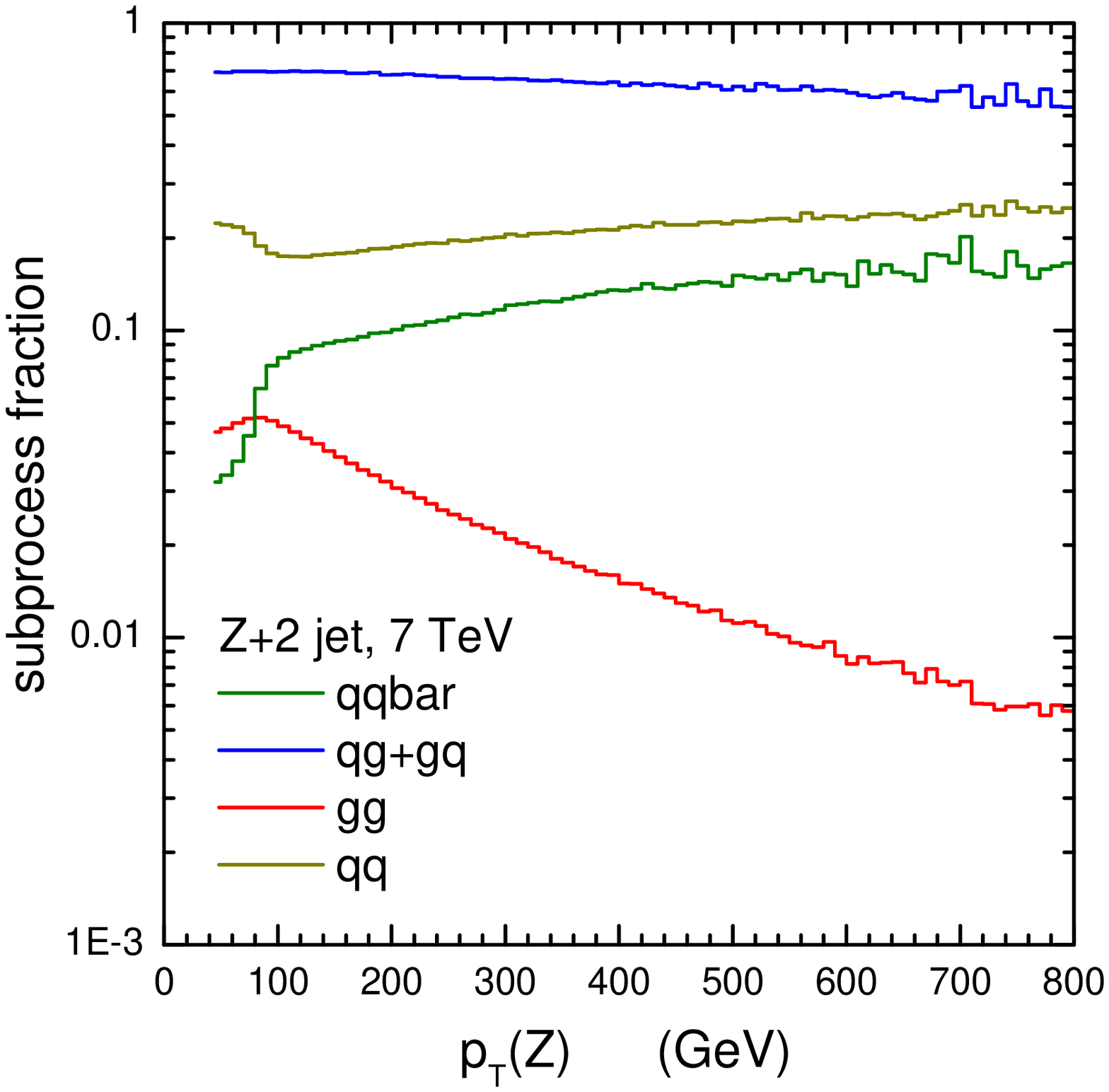}}
\caption{ \subref{fig:gr:dr} Ratio of the $Z+2$~jets and $\gamma+2$~jets $p_T(V)$ distributions 
at 7~TeV, for different values of the $\Delta R(V,j)> \Delta R_{\rm min}$ isolation cut.  
\subref{fig:gr:comp} Breakdown of the $Z + 2$~jet \gamb\ cross section into the different subprocess 
contributions defined in the text.}
\label{fig:gambosZgptratio2J}
\end{center}
\end{figure}

For the $V + 2$~jets production cross sections there is an additional complication 
in that the high--$p_T$ photon can be emitted collinearly to a high--$p_T$ quark, 
with the transverse momentum of the pair being balanced by an `away side' quark or 
gluon. For massless photons, quarks and gluons the matrix element is singular in 
this configuration and so the closer the photon is allowed to approach the quark, 
the smaller the $Z/\gamma$ ratio becomes, since there is no such collinear 
singularity for $Z$ production. To regulate the singularity, we impose a $\Delta R(V,j)> \Delta R_{\rm min}$ 
isolation cut, where $V=\gamma, Z$ and $j=q,g$.\footnote{Note that the requirement 
of photon isolation becomes more subtle beyond leading order in pQCD, since in this 
case the partonic jets can have non--zero \lq width' and the choice of jet algorithm 
influences the analysis. This issue is addressed in detail in Ref.~\cite{Bern:2011pa}. 
In our case we will be studying photon isolation for the full \pyt\ event simulation 
including hadronisation and experimental cuts. Note also that we neglect contributions 
to the $\gamma$ cross section involving photon fragmentation functions, \ie $f^{q\to\gamma}(z,Q^2)$. 
With the strong isolation requirements and high transverse momentum values used in 
our study, we expect such contributions to be small.} We also, of course, need to 
impose rapidity, transverse momentum and jet--jet separation cuts on the quark and 
gluons jets. For illustration, we choose $p_T(j) > 40$~GeV, $\vert y(j)\vert < 2.5$ 
(as for the 1--jet study above), and  $\Delta R(j,j)> 0.4$ to represent `typical' 
experimental cuts. 
Figure~\ref{fig:gr:dr} shows the ratio of $V + 2$~jet cross sections, as calculated 
using \gamb,  for different values  $\Delta R_{\rm min} = 0.05,0.1,0.2,0.4,0.6$ 
at 7~TeV. The ratio at high $p_T$ shows the expected dependence on the minimum 
separation. Note also that the ratio becomes insensitive to the isolation cut when 
the minimum separation becomes large, since far from the singularity the $Z$ and 
$\gamma$ phase space are affected more or less equally. A similar dependence on  
$\Delta R_{\rm min}$ is observed for the $V + 3$~jet ratios. From now on we take 
$\Delta R_{\rm min} = 0.4$ as our default choice and attribute a $\pm 5$\% uncertainty of 
these results based on the difference with respect to $\Delta R_{\rm min} = 0.6$.

The breakdown of the $Z + 2$~jet \gamb\ cross section at 7~TeV into the different 
subprocess contributions is shown in figure~\ref{fig:gr:comp}. We define these to be 
$q\bar q\to Vgg$ scattering (\eg figure~\ref{fig:diags}(c)), $qg,gq\to Vqg$  
(\eg figure~\ref{fig:diags}(d)), $gg\to Vq \bar q$ (\eg figure~\ref{fig:diags}(f)) 
and $qq\to Vqq$ (\eg figure~\ref{fig:diags}(e)), where a sum over quarks and antiquarks 
is implied. Note that quark--gluon scattering is by far the most dominant in the 
kinematic region studied here, and its fractional contribution is roughly independent 
of $p_T(V)$. The results also show that the second largest contribution in the 
2--jets case comes from the $qq$ subprocess, which approximately amounts to 20\%. 
This is in contrast to the 1--jet case, which is dominated by $qg$ and 
$q\bar{q}$ scattering. The corresponding subprocess breakdown of the $\gamma + 2$~jet cross 
section is similar.

\begin{figure}[ht]
\begin{center}
\subfigure[][]{\label{fig:gambosZgptratio1J2J3J} \includegraphics[width=0.48\textwidth]{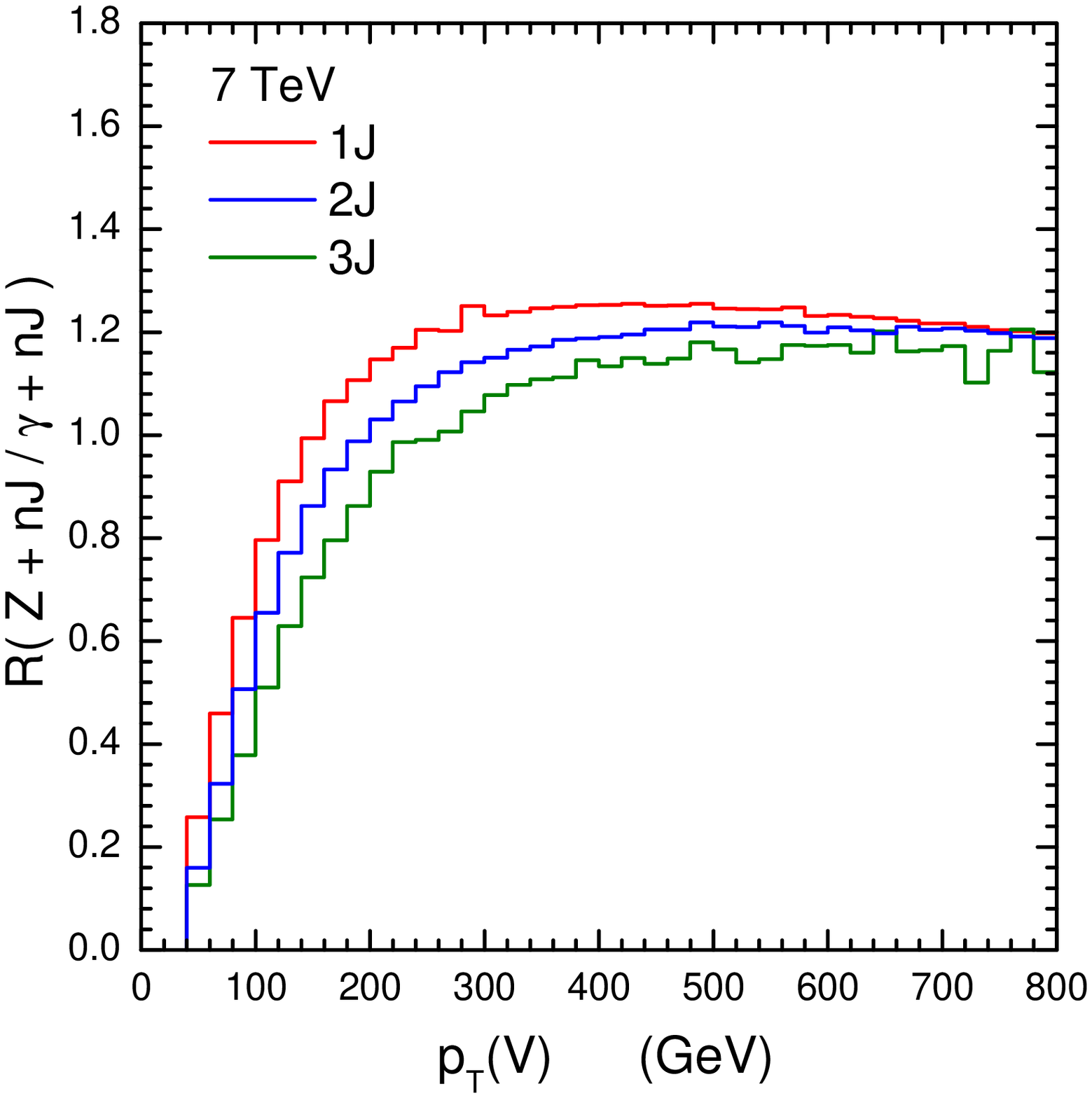}}
\subfigure[][]{\label{fig:pyt:mj} \includegraphics[width=0.48\textwidth]{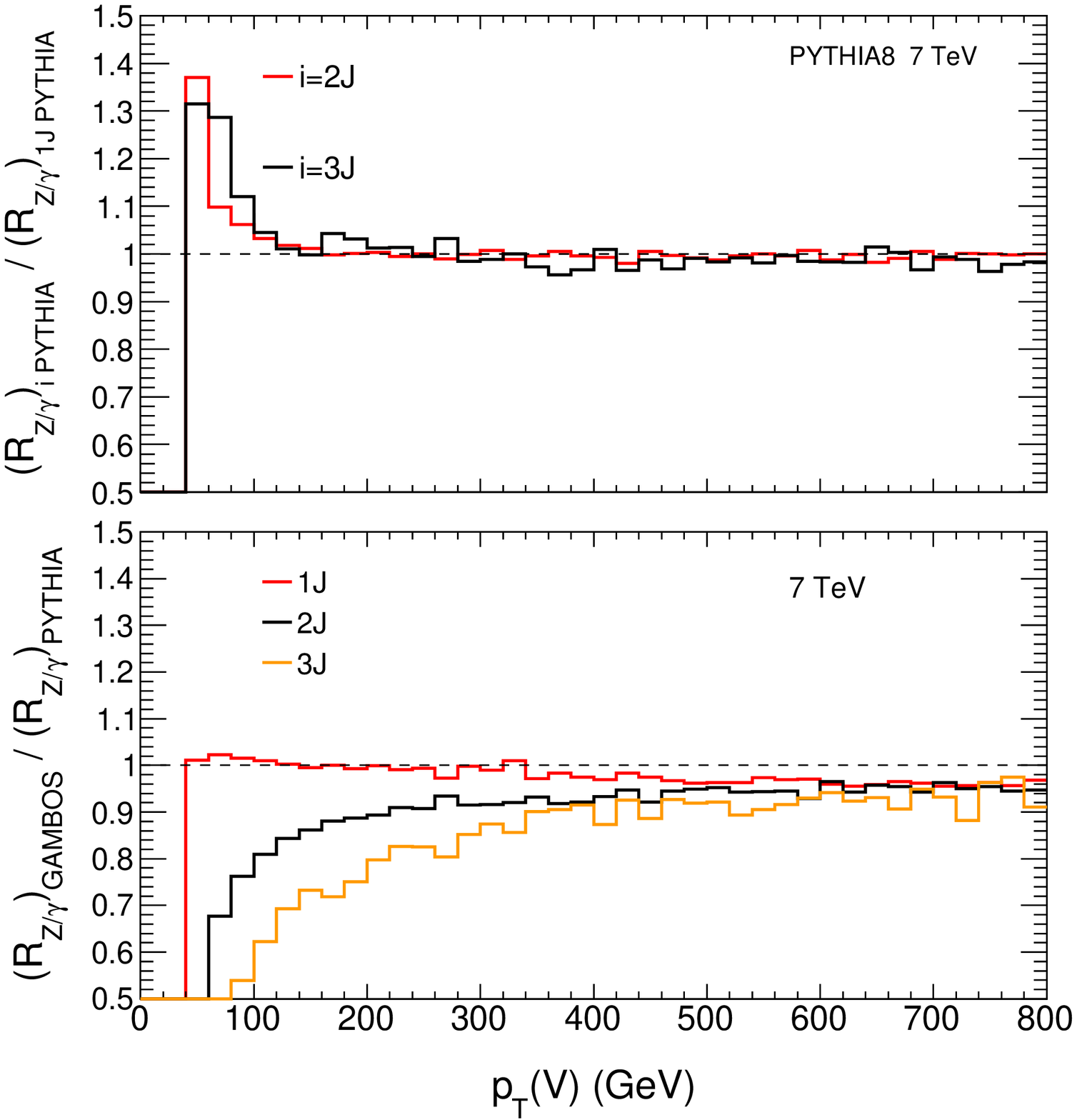}}
\caption{\subref{fig:gambosZgptratio1J2J3J} $Z/\gamma +1,2,3$~jet \gamb\ cross 
section ratios as a function of $p_T(V)$, with the isolation cut $\Delta R_{\rm min} = 0.4$ . 
\subref{fig:pyt:mj} Effect from multijet requirement on the \pyt\ ratio (upper) 
and difference with respect to \gamb\ (lower).
}
\end{center}
\end{figure}

In figure~\ref{fig:gambosZgptratio1J2J3J} we show the $Z/\gamma +1,2,3$~jet
\gamb\ cross section ratios as a function of $p_T(V)$, with $\Delta R_{\rm
min} = 0.4$ and other cuts as before. We see that the 2,3 jet ratios are
slightly smaller than the 1~jet ratio at moderate and high $p_T$. The small
difference arises from three effects: (i) the dependence of the 2,3~jet
cross sections on $\Delta R_{\rm min} = 0.4$, (ii) the additional $qq$ and
$gg$ scattering diagrams, the net effect of which is to decrease the 2,3
jet ratio, as already explained in section~\ref{sec:gambos}, and (iii) the
fact that for a fixed $p_T$, increasing the number of jets increases the
overall invariant mass of the final--state system, and also therefore the
values of the parton momentum fractions. This in turn decreases the $d/u$
ratio, and also the $Z/\gamma$ ratio, see figure~\ref{fig:ud:ud}.

For more than one jet, the additional jets in the \pyt\ simulation are produced 
by parton showering. In order to characterise the difference with respect to exact 
MEs, the following events were used for the 2-- and 3--jet results from \pyt\ at 
the parton level, where the aim is to produce events with a similar parton level 
topology to the \gamb\ events based on the multijet MEs. In the 2--jet case, events 
were generated with both initial (ISR) and final state radiation (FSR) enabled and 
only those with exactly 2 jets within the allowed $p_T$ and $y$ acceptance were 
considered. These jets either correspond to the parton from the hard scatter together 
with an ISR emission or from events with no accepted ISR emission, but where the 
hard parton, due to FSR, branches into two partons within the allowed acceptance. 
In addition, only events with the required $V -$ jet and jet--jet $\Delta R$ 
separation were considered. The same event selection was used for the parton level 
3--jet case, where the three jets either correspond to the hard parton together with 
two ISR emissions, or a FSR branch of the hard parton together with one ISR emission, 
or two FSR emissions. 
The multijet ratios from \pyt\ were found to be very similar to the one jet case. 
This is illustrated in the upper plot of figure~\ref{fig:pyt:mj}, which shows the 
\pyt\ $R_{2{\rm jet}}/R_{1{\rm jet}}$ and $R_{3{\rm jet}}/R_{1{\rm jet}}$ ratios.\footnote{The 
large deviation in the first bins is an artefact of the kinematic acceptance. In 
the 1--jet case, events will populate these bins according to the normal underlying 
distribution. However, since shower emissions can only occur with $p_T$ smaller 
than the hard process, the 2,3 jet events will be peaked toward the upper bin edge. 
Due to the sharp rise of the ratio at low $p_T$, the average ratio value in these 
bins will therefore be significantly different for the higher jet multiplicities.}
%
The corresponding \gamb\ $Z/\gamma$ ratios are therefore slightly smaller as shown 
in the lower plot of figure~\ref{fig:pyt:mj}.
These results illustrate the difference between the two individual approaches in 
the multijet case as well as the importance of using a multiparton ME based program 
in the actual analysis, where the precision related to the ME calculation was discussed 
above in connection with figure~\ref{fig:gambosZgptratio2J}.
By comparing the \gamb\ and \pyt\ ratios at high $p_T$, we can extract correction 
factors for the \pyt\ ratios to take account of the missing contributions, however, as 
seen in figure~\ref{fig:pyt:mj} these correction factors are not large.

\section{Full event simulation}

The ability of \pyt\ to simulate full events was used both to investigate the 
robustness of the $Z/\gamma$ cross section ratio as well as its potential use 
in estimating the \zvv background in searches for new physics at the LHC. For 
simplicity, the experimental aspects related to the photon analysis attempt to 
follow as closely as possible what is commonly used in ATLAS analyses \cite{atlas:gma,atlas:gmb}, 
and for the new physics scenario we focus on the SUSY zero--lepton search \cite{atlas:susya,atlas:susyb}, 
where SM \zvv production is one of the main backgrounds. Due to the phenomenological 
nature of this study, we neglect any experimental photon inefficiencies, apart 
from the isolation criteria discussed below, as well as any backgrounds (\eg 
$\pi^0 \to 2\gamma$), which in any case are expected to be relatively small in 
the high $p_T$ region of interest.

\subsection{Effects on the ratio \label{sec:recorat}}

The same \pyt\ processes were used as in the parton--level study, but with the 
full parton shower, hadronisation, multiple interaction and particle decay simulation 
enabled. The default settings of v8.150 were used, for which general performance 
results can be found in \cite{Buc:11}. 
The same selection was used as for the 1--jet parton--level results, here corresponding 
to an inclusive jet selection.
The main differences with respect to the parton--level results come from using the 
final state boson momentum as well as from using jets reconstructed from the final 
state particles, rather than being represented by single partons. The jets were 
reconstructed using the {\sc FastJet} library \cite{Cac:06,Cac:08,Cac:web} and  
were based on all final--state particles except leptons and any photons with 
$p_T^\gamma > 30$~GeV. The anti--$k_t$ algorithm was used with a $R$ parameter of 
0.4, also in accordance with the ATLAS analysis. In the following, these results 
from full event simulation are referred to as obtained at particle level.

\begin{figure}[ht]
\begin{center}
\subfigure[][]{\label{fig:fs:iso} \includegraphics[width=0.48\textwidth]{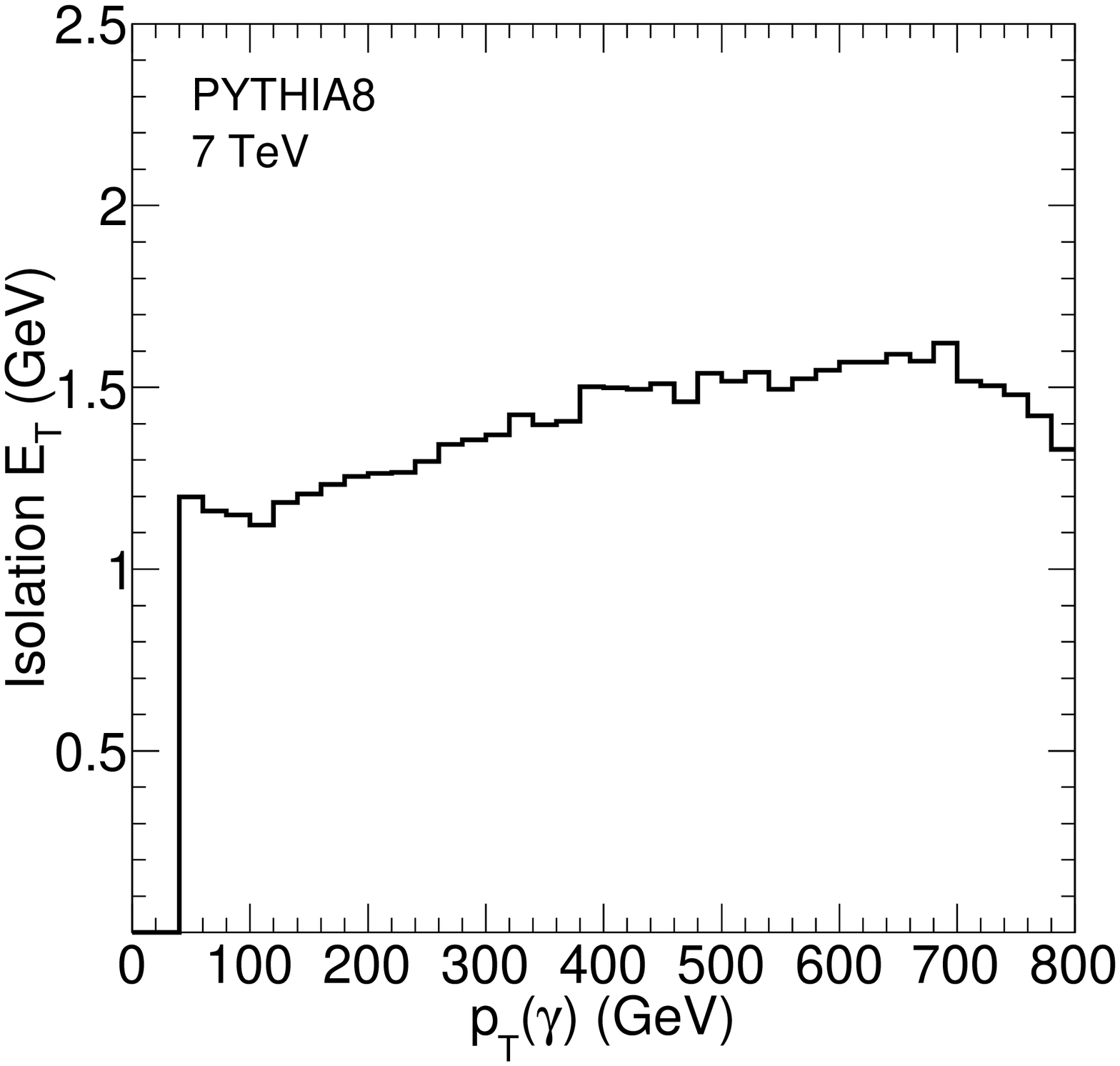}}
\subfigure[][]{\label{fig:fs:rat} \includegraphics[width=0.48\textwidth]{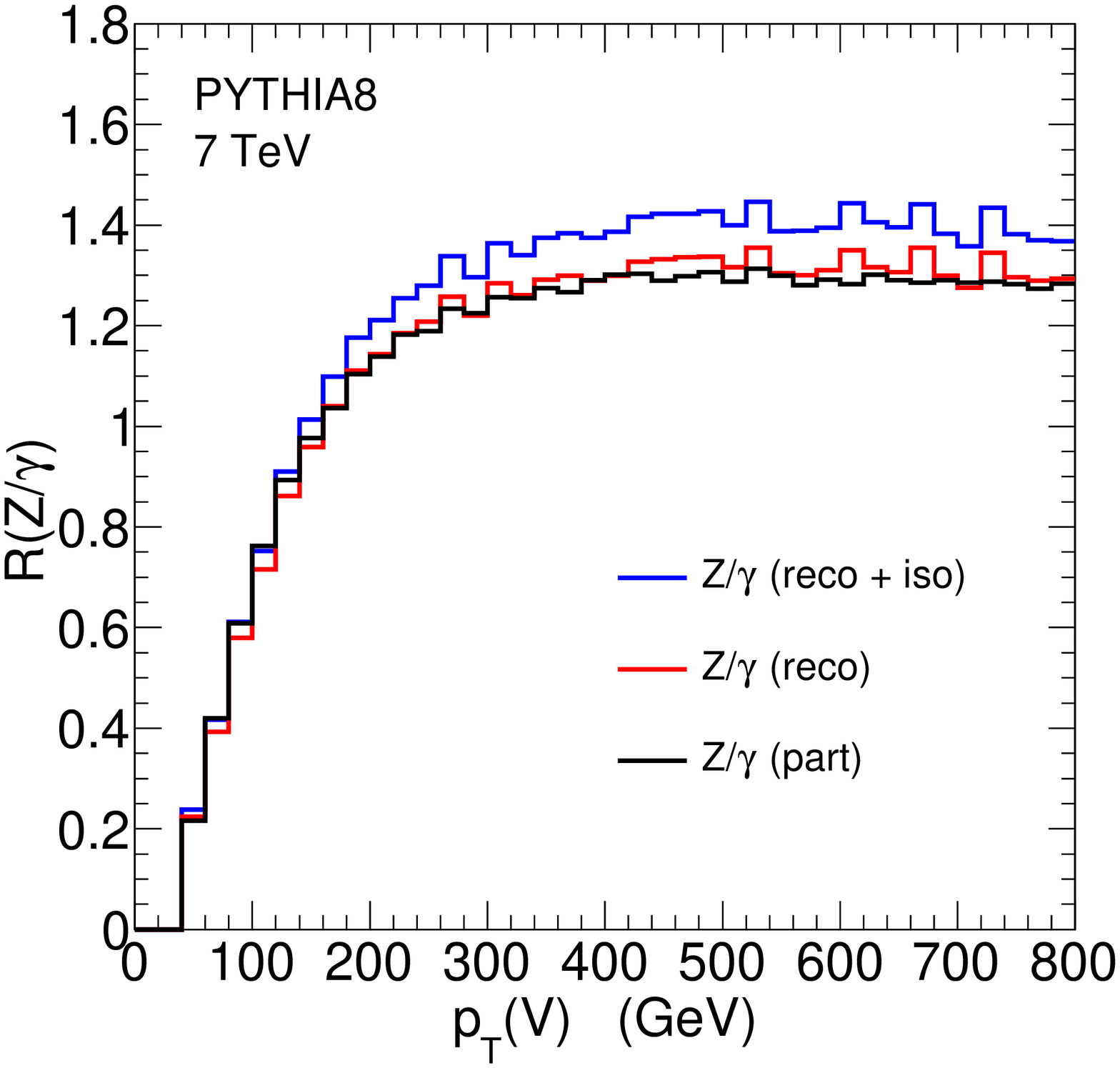}}
\caption{ \subref{fig:fs:iso} Transverse energy inside the photon isolation cone 
as a function of the photon $p_T$. \subref{fig:fs:rat} $Z/\gamma$ cross section 
ratio as a function of the photon $p_T$. Parton (part) and particle level results 
(reco) are shown as well as for events passing the isolation criteria (reco + iso).
\label{fig:gmisorat}}
\end{center}
\end{figure}

The experimental analysis in ATLAS uses a photon isolation criterion in order 
to suppress QCD background and this quantity was found to be well described at 
MC generator level. This isolation criterion requires the transverse energy 
within a $\Delta R [ =  \sqrt{\Delta\eta^2 + \Delta\phi^2} ] = 0.4$ cone around 
the photon ($E_T^{iso}$) not to exceed 4~GeV. Since the anticipated use of the 
ratio here is to estimate the number of $Z$ events from measured $\gamma$ events, 
the effect from this isolation requirement is also addressed. The fact that only 
isolated photons are considered implies an even stronger photon--jet separation 
than the one used above for the parton level results, ensuring an acceptance where 
\gamb\ provides robust calculations. The mean transverse energy within the photon 
isolation cone is shown in figure~\ref{fig:fs:iso} as a function of the photon $p_T$. 
This plot is based on events which pass the above selection applied to the final 
state photon as well as the reconstructed jets. A small increase with $p_T$ is 
shown, but with values well below 4~GeV over the whole range. In spite of the 
increasing hadronic recoil with larger boson $p_T$, a relatively constant 
inefficiency of about 5\% was found over the full $p_T$ range.

Figure~\ref{fig:fs:rat} shows the $Z/\gamma$ cross section ratio at parton 
(part) and particle level (reco) as well as at particle level where the isolation 
requirement is applied (reco+iso). 
%
%
Good agreement is evident between the results obtained at parton and particle level, 
where the difference is well below 5\% at high $p_T$, and the increase of the ratio 
due to the photon isolation criterion is of order 6\% at high $p_T$.

\subsection{Background estimate for a zero lepton SUSY search}

\begin{figure}[ht]
\begin{center}
\subfigure[][]{\label{fig:fs:ptj2} \includegraphics[width=0.48\textwidth]{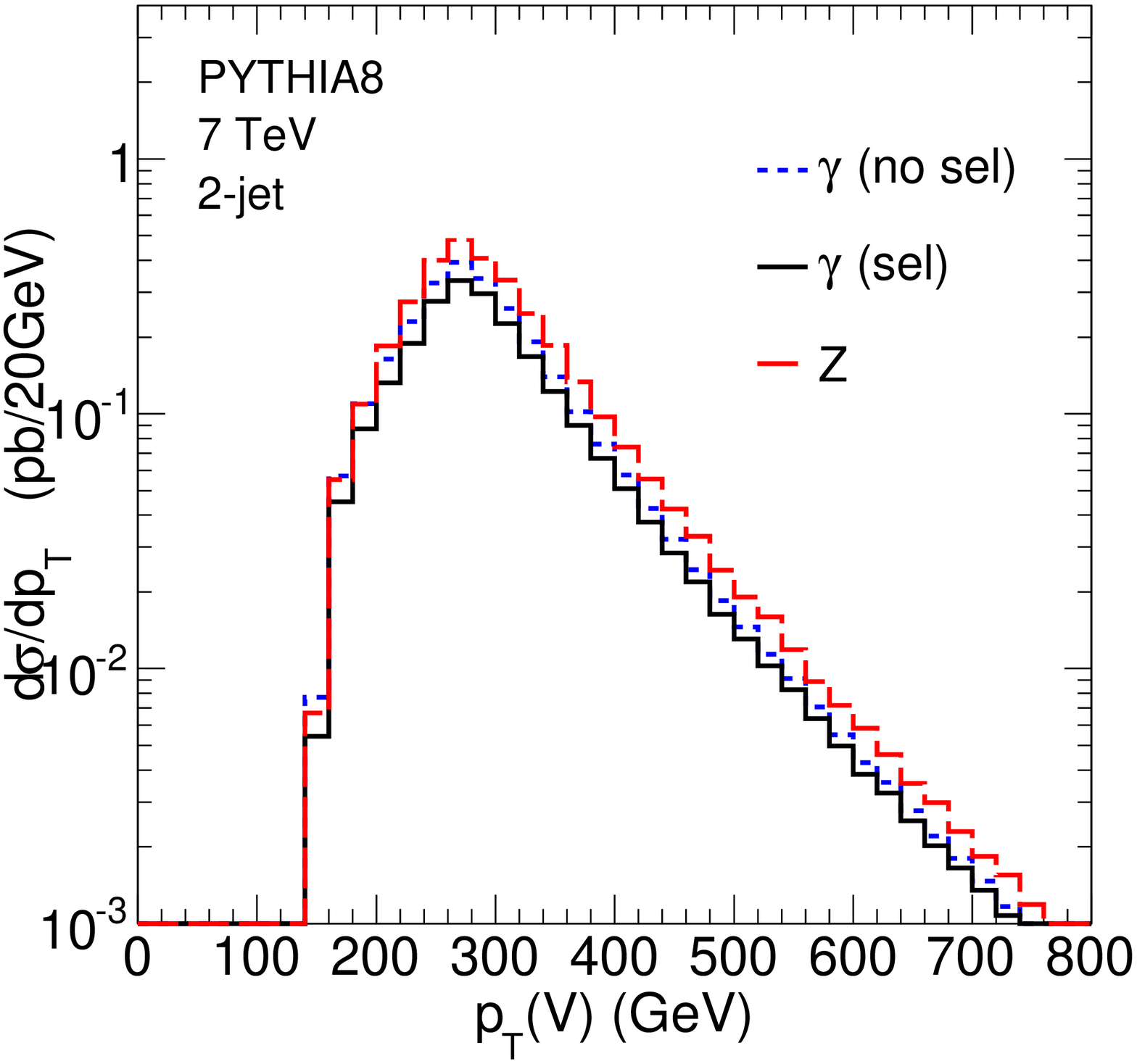}}
\subfigure[][]{\label{fig:fs:etaj2} \includegraphics[width=0.48\textwidth]{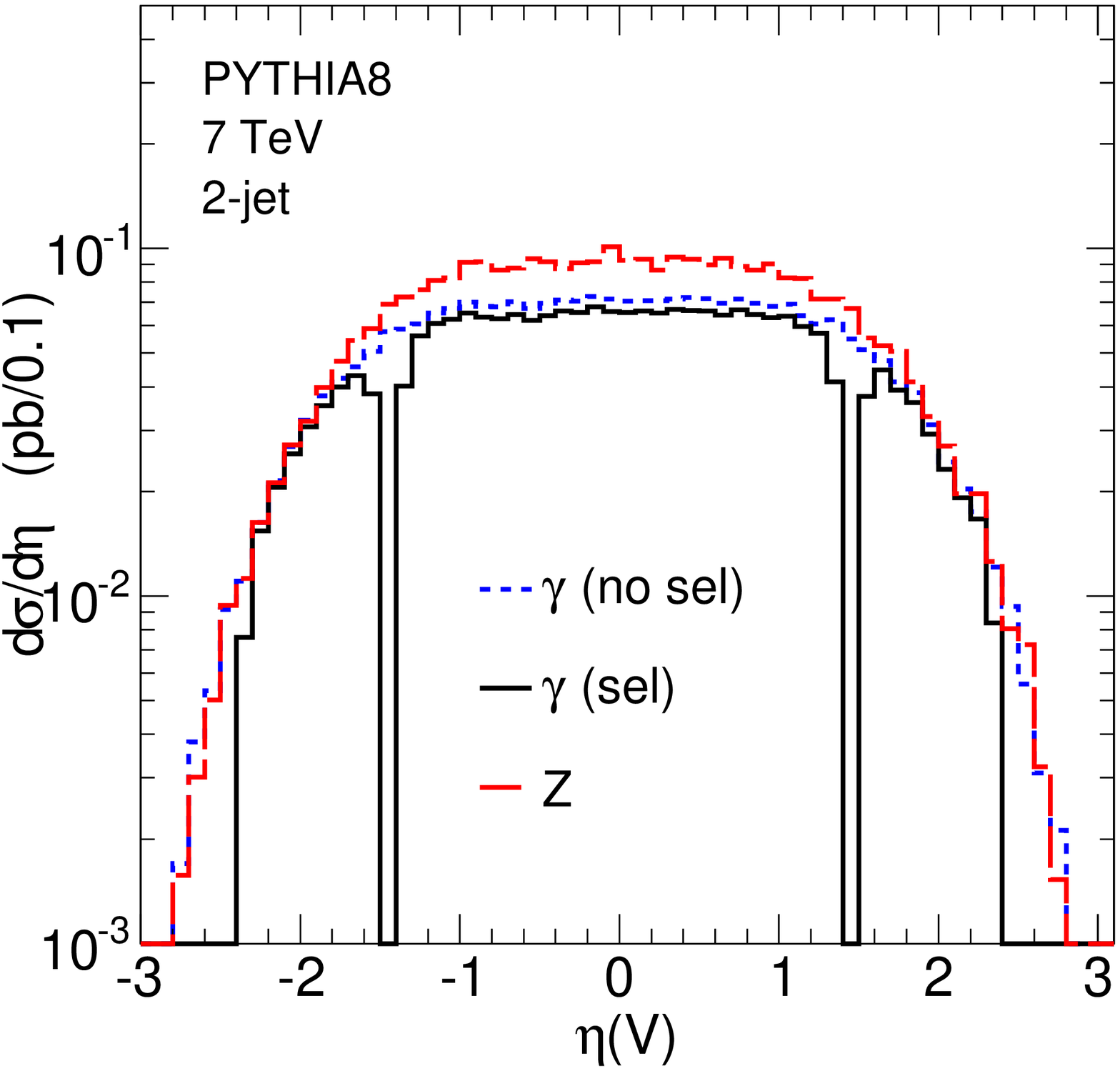}}
\caption{\subref{fig:fs:ptj2} Boson $p_T$  and  \subref{fig:fs:etaj2} $\eta$ 
distributions from events passing the 2--jet SUSY selection. Results from 
photon events both with (sel) and without (no sel) applying the photon 
analysis selection are shown together with results from $Z$ events (Z).
\label{fig:susy2pteta}}
\end{center}
\end{figure}

\begin{figure}[ht]
\begin{center}
\subfigure[][]{\label{fig:fs:ptj3} \includegraphics[width=0.48\textwidth]{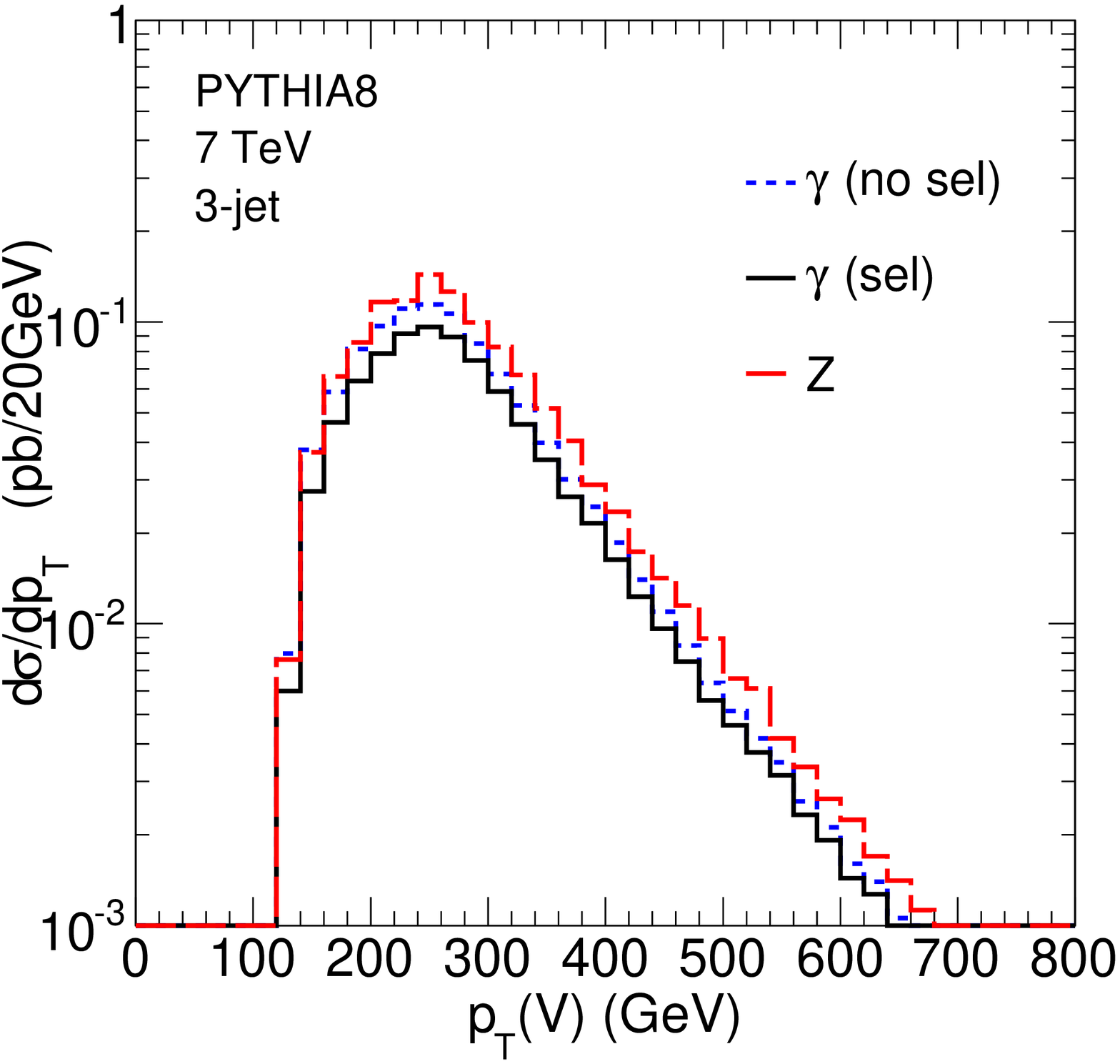}}
\subfigure[][]{\label{fig:fs:etaj3} \includegraphics[width=0.48\textwidth]{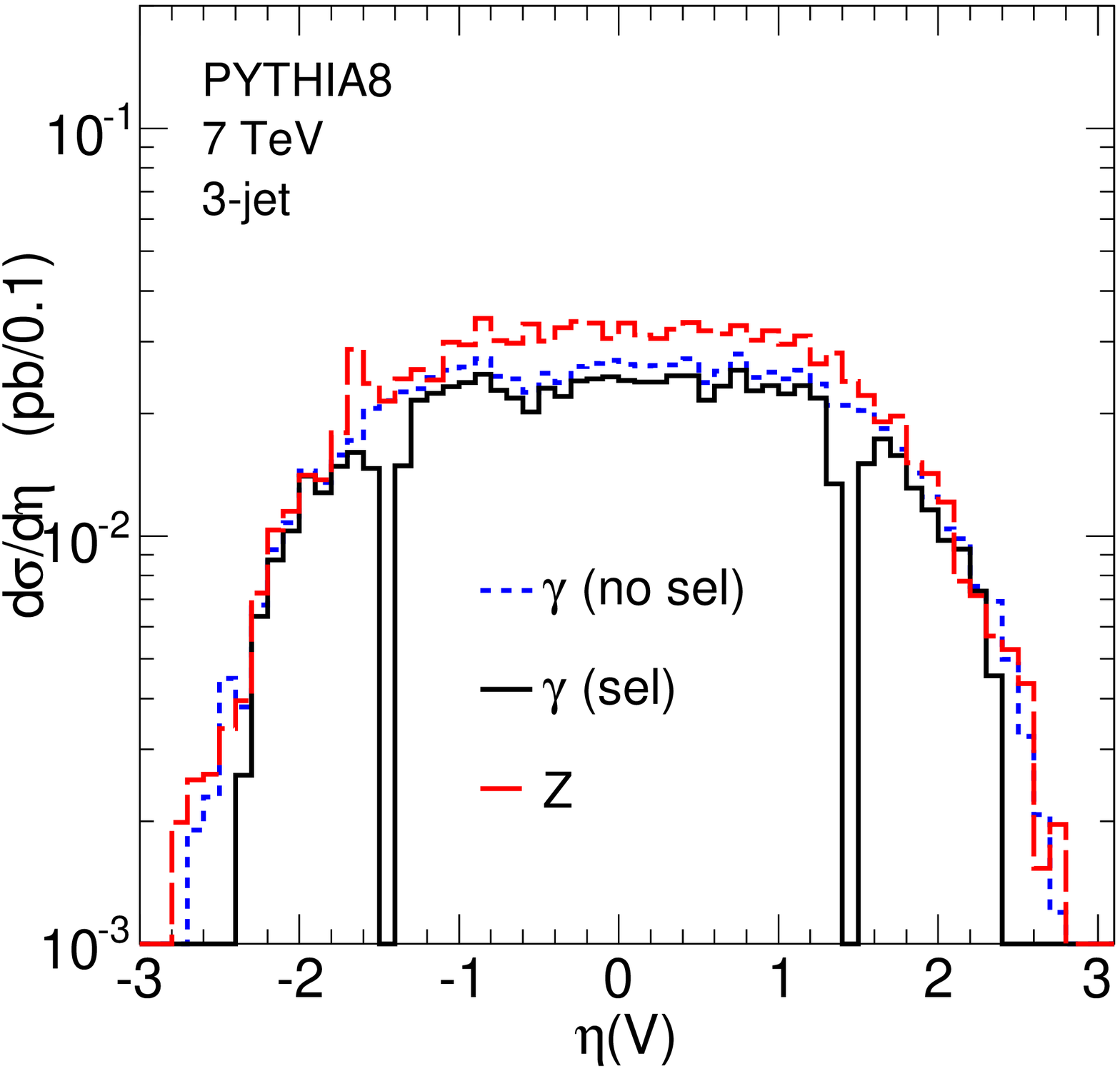}}
\caption{ \subref{fig:fs:ptj3}  Boson $p_T$ and \subref{fig:fs:etaj3} $\eta$  
distributions from events passing the 3--jet SUSY selection. Results from 
photon events both with (sel) and without (no sel) applying the photon 
analysis selection are shown together with results from $Z$ events (Z).
\label{fig:susy3pteta}}
\end{center}
\end{figure}

This section demonstrates the estimation of \zvv background for a zero lepton 
SUSY search using photon events. This is done using the \pyt\ results and is 
meant to serve as a general example, since the method could be used also for 
other new physics searches where \zvv production contributes with a significant 
background. The method involves the following steps:
\begin{itemize}
\item{Photon event selection. Select a photon event sample using a loose enough 
selection, with respect to the photon and jets, to contain as many events as 
possible that will pass the final selection. This is represented here by the 
criteria $p_T(\gamma) > 45$ GeV, $|\eta(\gamma)| < 2.37$, excluding $1.37 < |\eta(\gamma)| < 1.52$ 
and $E_T^{iso} < 4$ GeV, based on the ATLAS photon analysis~\cite{atlas:gma,atlas:gmb}.}
\item{SUSY event selection. Apply the SUSY selection to the photon events, 
where the photon $p_T$ represents the missing transverse energy from the $Z$ 
in the events to be estimated. A 2--jet as well as 3--jet SUSY selection is 
used, based on the ATLAS search. 2--jet (3--jet): $p_T(j_1) > 120$ GeV, 
$p_T(j_2) > 40$ GeV, ($p_T(j_3) > 40$ GeV), $|\eta(j_i)| < 2.5$, $p_T(V) > 100$ GeV, $\Delta \phi(V,j_i) > 0.4$, 
$p_T(V) / m_{eff} > 0.3$ and $m_{eff} > 500$ GeV. Here $j_i$ represents the $i^{th}$ 
leading jet and $m_{eff}$ is the SUSY discriminating variable used in~\cite{atlas:susya,atlas:susyb}, 
defined as the scalar  sum of the $p_T$ from the jets and the boson in the event.}
\item{Subtract backgrounds and correct for experimental efficiencies. This is 
represented here only by the isolation efficiency.}
\item{Convert photon events, inside the acceptance of the analysis, to $Z_{\nu \nu}$ 
events using the cross section ratio, $R(p_T^V) \cdot Br(Z \rightarrow \nu \nu)$. 
As discussed in the previous sections, in an analysis of real LHC data, $R(p_T^V)$ 
should be based on results from exact multijet MEs and using an appropriate 
jet selection.}
\item{Correct for acceptance constraints implied by the photon analysis, \eg the 
$\eta(\gamma)$ selection criteria.}
\end{itemize}
The main intention with this method is that all necessary corrections as well as 
theoretical input are related to the vector bosons, whereas all requirements with 
respect to the experimentally more challenging reconstructed jets, are identical 
for the $Z$ and $\gamma$ events. As shown in the previous section, the ratio is 
affected by requiring jets. However, since this effect is a consequence of changing 
the mixture of couplings imposed by the relevant initial partons together with 
their PDFs, it is small even for drastically different jet criteria and should be 
yet smaller with respect to experimental jet uncertainties, such as energy scale 
and resolution.

The photon event selection imposes some unavoidable criteria which are not experienced 
by the $Z_{\nu \nu}$ background and therefore has to be corrected for. The implications 
of this selection on the final sample are, however, relatively mild for the following 
reasons. The photon $p_T$ requirement is significantly softer than the subsequent 
selection. A photon isolation criterion, of some kind, is required in order to obtain 
accurate calculations of the ratio. However, this is often also used in the SUSY 
selection for more experimental reasons, such as to prevent fake missing transverse 
energy caused by a high $p_T$ jet, \ie the $\Delta \phi(V,j_i)$ requirement above. 
In addition, the fact that the $Z$ and $\gamma$ processes have the same phase space 
when $p_T \gg M_Z$ means that at high $p_T$, the $\eta$ distributions from the 
$\gamma$ and $Z$ events converge toward the same distribution, which becomes increasingly 
central with higher boson $p_T$. Therefore any acceptance corrections will become the 
same for the $Z$ as for the $\gamma$ events.

The SUSY selection, based on the $p_T$ of the boson and jets in the event, is then 
identical for the two event types and should not require any related corrections. 
Due to the fact that the bosons are recoiling against the hadrons, the SUSY selection 
will in principle act as a non--trivial high boson $p_T$ criteria. For this reason 
the $\gamma$ events passing the SUSY selection can be converted into $Z$ events based 
only on the boson kinematics. Again due to the convergence of the $Z$ and $\gamma$ 
phase space at high $p_T$, the cross section ratio becomes insensitive to the particular 
$\eta(V)$ criteria used and is hence determined by the $p_T(V)$. The precision of 
this method is therefore mainly related to the photon analysis part, which is expected 
to be precise at high photon $p_T$, and the theoretical knowledge of the $Z/\gamma$ 
cross section ratio.

\begin{figure}[ht]
\begin{center}
\subfigure[][]{\label{fig:fs:zgj2} \includegraphics[width=0.48\textwidth]{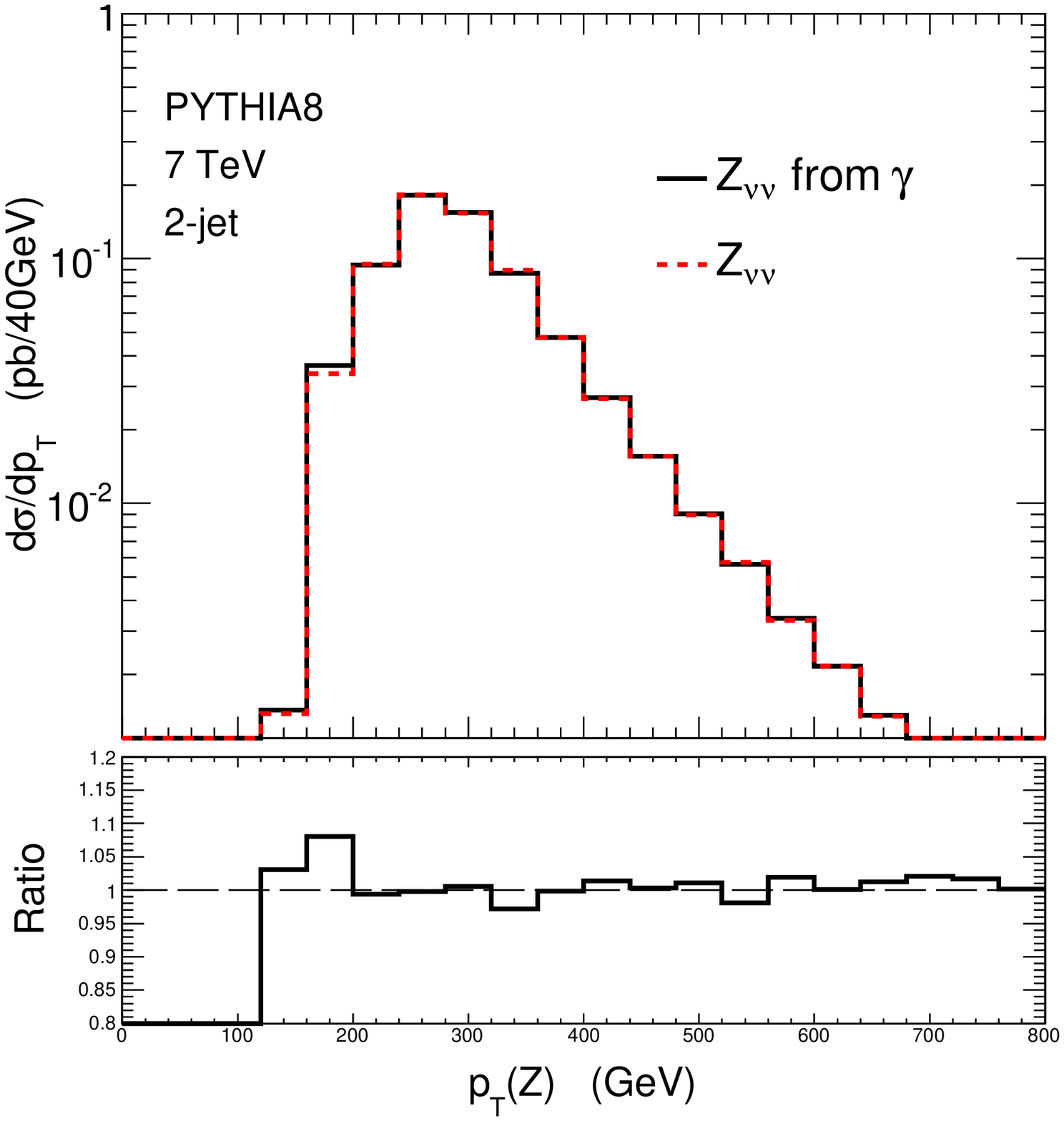}}
\subfigure[][]{\label{fig:fs:zgj3} \includegraphics[width=0.48\textwidth]{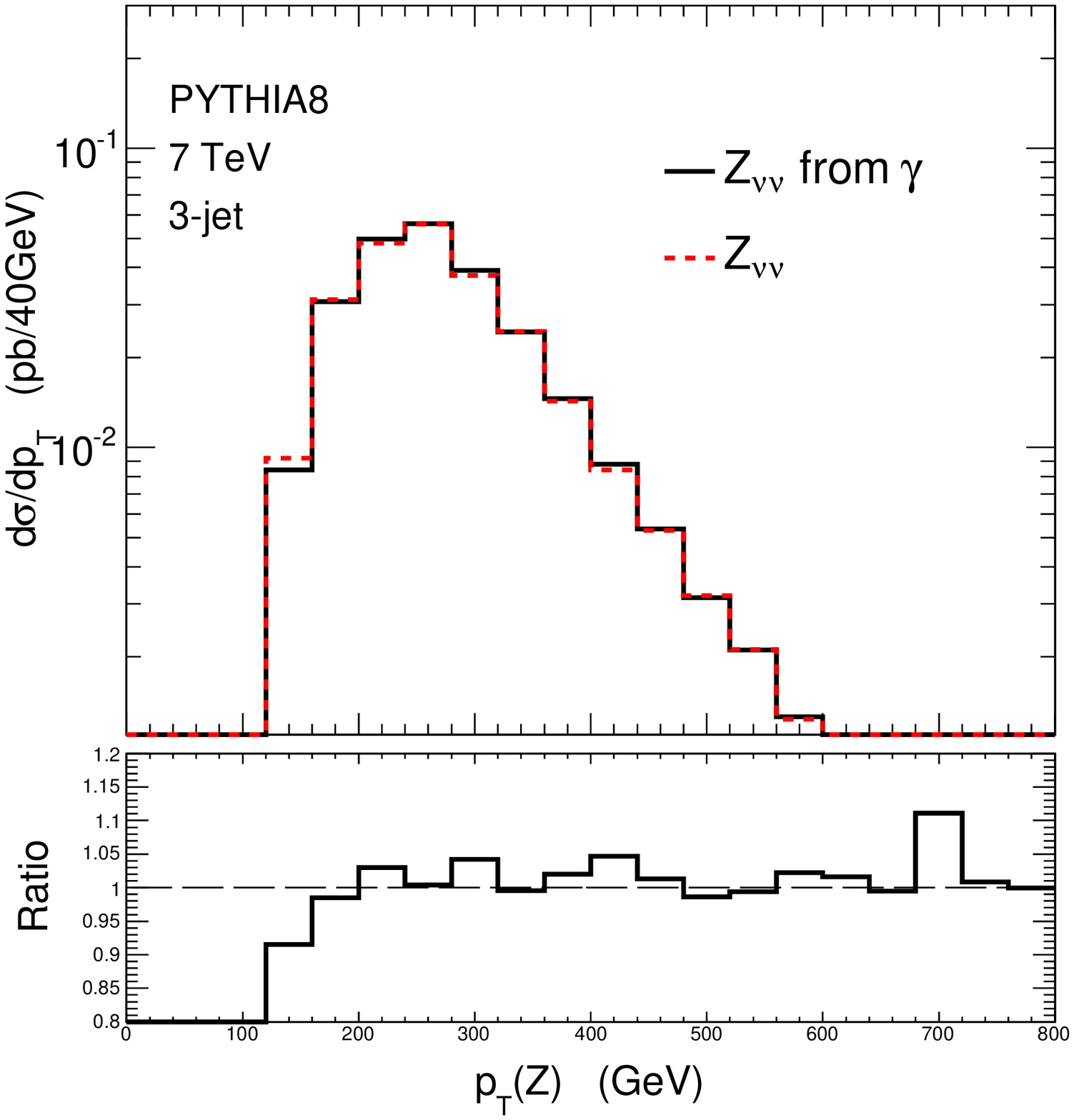}}
\caption{Differential cross section as a function of $p_T(Z)$ for $Z \rightarrow \nu \nu$ 
events passing the  \subref{fig:fs:zgj2} 2--jet and  \subref{fig:fs:zgj3} 3--jet SUSY selection. 
The predictions using $\gamma$ events ($Z_{\nu\nu}$ from $\gamma$) are compared to results 
from direct MC simulation of the \zvv process ($Z_{\nu\nu}$).
\label{fig:susygmz}}
\end{center}
\end{figure}

In figures~\ref{fig:susy2pteta} and \ref{fig:susy3pteta} the boson $p_T$ and $\eta$ 
distributions are shown after the 2--jet and 3--jet SUSY selections respectively. 
The distributions for photon events, with (sel) and without (no sel) passing the 
photon selection, as well as $Z$ events ($Z$) are shown. The $p_T$ distributions 
show that both the SUSY selections mainly select events with a boson $p_T$ in the 
range 250 to 300 GeV. The $\eta$ distributions also show the similarity in shape 
of the distributions from the $Z$ and $\gamma$ processes. Both the $p_T$ and $\eta$ 
distributions show that the effect from the photon selection, \ie acceptance and 
isolation requirements, is relatively small also after the SUSY selection and the 
difference between the $Z$ and $\gamma$ results is consistent with the cross section 
ratio at relevant boson $p_T$ values. In the $\eta$ distributions, the difference 
from applying the photon selection reflects the impact of the isolation criteria 
alone and it is shown that the additional jet requirements from the SUSY selections 
do not change the isolation efficiency dramatically. The results from the 2--jet 
and 3--jet selections do have slight differences, but the overall characteristics 
discussed are the same.
Figure~\ref{fig:susygmz} shows the $Z$ $p_T$ distributions from $\gamma$ events 
passing the two SUSY selections, converted into \zvv according to the method 
outlined above, together with the results obtained from simulating \zvv directly.
Since the \pyt\ ratios do not show any jet multiplicity dependence, the ratio shown 
in figure~\ref{fig:gmisorat} (reco) was used in this analysis example for both the 
2--jet and 3--jet results. 
As expected, the two distributions agree within the statistical uncertainty of 
the simulation, which is smaller than 5\% in the bulk of the distributions.

The uncertainties on the final background estimate, related to the experimental 
aspects of this analysis, are not covered by this study. However, as discussed 
above, the design of the method should limit the exposure mainly to the corrections 
from the photon event selection, which are expected to be precise at high boson 
$p_T$. In addition, as discussed in section~\ref{sec:recorat}, the impact associated 
with full event simulation on the cross section ratio was found to be small. 
As shown in figure~\ref{fig:fs:rat}, the ratio increases, by about 6\% in the high 
$p_T$ region, when including the isolation efficiency, whereas a significantly 
smaller effect is seen when going from parton to particle level.
The final uncertainties on $R(Z/\gamma)$ from such effects are therefore expected 
to be at the percent level. 
A theoretical uncertainty, here with respect to the hard process calculations behind 
$R(Z/\gamma) \cdot Br(Z\rightarrow \nu\nu)$, that significantly exceeds 10\% is 
therefore likely to dominate the uncertainty of the method.

Table~\ref{tab:errors} presents the overall $R \cdot Br$ values obtained for the results 
shown in figure~\ref{fig:susygmz}. This includes both the value from \pyt\ ($R^{P}$), 
that was actually used in the plot, as well as the corrected value ($R^{ME}$) based 
on the ME results. The table also includes the uncertainties associated with the ME 
calculation ($\varepsilon_{ME}$), the scales ($\varepsilon_{\mu}$) and the PDFs 
($\varepsilon_{PDF}$), see section~\ref{sec:partonlevel}. 
The results show a total uncertainty of $\pm 7$\% and indicate that 
the theoretical uncertainty for results obtained by an appropriately configured MC 
program, which uses ME amplitudes for the hard jets, should be within 10\%. The SUSY 
selection used in the experimental analysis will evolve with an increasing amount of 
LHC data. However, such an evolution is expected to effectively imply a harder boson 
$p_T$ requirement and since the uncertainties in table \ref{tab:errors} are valid for $p_T(V) > 100$ 
GeV, the same conclusions should hold also for harder selections.\footnote{At least up to 
$p_T(V) = 800$ GeV, which is the maximum value included in this study.}

\begin{table}
\begin{center}
  \begin{tabular}{|l|c|c|c|c|c|c|}
    \hline
    Selection 
    & $R^{P} \cdot Br$ 
    & $R^{ME} \cdot Br$ 
    & $\varepsilon_{ME}$ 
    & $\varepsilon_{\mu}$ 
    & $\varepsilon_{PDF}$  
    & $\varepsilon_{Tot}$ \\ 
    \hline 
    \hline
    2--jet  & 0.254 & 0.234 & 5\% & 3\% & 4\% & 7\% \\ \hline
    3--jet  & 0.246 & 0.207 & 5\% & 3\% & 4\% & 7\% \\ \hline
  \end{tabular}
  \caption{The overall $R \cdot Br$ values obtained after the SUSY selections, directly from 
    \pyt\ ($R^{P}$) as well as corrected values with respect to the ME results ($R^{ME}$). The 
    uncertainties associated with the ME calculation ($\varepsilon_{ME}$), the scales 
    ($\varepsilon_{\mu}$) and the PDFs ($\varepsilon_{PDF}$) are also shown.}
  \label{tab:errors}
\end{center}
\end{table}


\section{Summary and conclusions}

One of the best methods to calibrate the irreducible background from $Z(\rightarrow \nu \nu)$+jets, 
to beyond the SM searches at the LHC, comes from using $\gamma$+jets data. The method 
utilises the fact that at high boson $p_T$ ($\gg M_Z$) the event kinematics converge 
for the two processes and the cross sections differ mainly due to the boson couplings. 
The advantage comes from large statistics, compared to alternative methods using 
$Z(\rightarrow ee,\mu\mu)$+jets events, together with the clean signature, with respect 
to experimental efficiencies and background, at high photon $p_T$. Hence, a precise 
prediction from theory of the $Z/\gamma$ cross section ratio, $R(Z/\gamma)$, is required. 
The similarity between the two processes should allow for a robust prediction of 
$R(Z/\gamma)$, given careful attention to the modelling of the jets.  

The general dependence of $R(Z/\gamma)$ on the mixture of boson couplings, which is 
determined by the initial state partons of the relevant amplitudes and their corresponding 
PDFs, has been illustrated. Relatively accurate values can be obtained even using rough 
approximations in the 1--jet case, whereas a larger set of amplitudes becomes necessary 
when 2 or more jets are required. The ratios have been studied at parton level using both 
the (LO) \pyt\ as well as the (multijet ME) \gamb\ programs, which allows us to disentangle 
effects associated with the two approaches. The impact from exact MEs when requiring different 
numbers of jets was found to be significant, but uncertainties were found to be within 5\% for the 
acceptance from typical experimental cuts. The corresponding uncertainties related to the 
PDFs and scale choice were found to be less than 4\% and 3\% respectively.

The \pyt\ MC program was used to investigate effects on $R(Z/\gamma)$ associated 
with full event simulation as well as performing a proof of principle analysis example. The 
effects investigated were found to be small and indicate that a theoretical precision\footnote{Again, 
referring to the hard QCD process calculations behind $R \cdot Br$.} at the 10\% level is 
required, in order not to significantly degrade the performance of the method. The total 
theoretical uncertainty was found to be 7\%, indicating that the results obtained by MC 
simulations, including exact multijet MEs, should be within the 10\% level. These results 
should also hold for similar 2-- and 3--jet selections, given  that the effective $p_T$ 
requirement on the boson is harder than what is used in the example analysis. Note that all 
our theoretical cross sections are evaluated in leading--order pQCD. It will be important 
to check, using for example the techniques of \cite{Bern:2011pa}, that the ratio predictions 
are indeed stable -- at least to the required accuracy -- with respect to higher--order pQCD corrections. 

Finally, one type of correction that has not been included in our theoretical study is
high-order electroweak corrections. Although these are intrinsically small, and many 
will again cancel in the $Z/\gamma$ ratio, there is an important class of correction 
involving $W$ and $Z$ virtual exchanges that does not cancel in the ratio. The impact 
on both the $Z$ and $\gamma$ distributions have been studied in~\cite{Maina:2004rb,Kuhn:2005gv}.
It was shown that non-cancelling Sudakov-type logarithms $\sim \alpha \log^2(p_T(V)^2/m_W^2)$ 
appear at high $p_T(V)$, and decrease the $Z/\gamma+1$~jet ratio by $6\%$ ($11\%$) at 
$p_T(V) = 300\; (800)$~GeV~\cite{Kuhn:2005gv}. However care is needed in the interpretation 
of this result, since the emission of {\it real} $W$ bosons is expected to compensate 
the virtual Sudakov logarithms to some extent~\cite{Baur:2006sn,Bell:2010gi}. It is 
therefore important to carry out a full analysis of higher-order electroweak corrections 
for the multijet processes and acceptance cuts studied in this paper. 

\acknowledgments{This work was supported in part by the EU Marie Curie Research Training Network ``MCnet'', 
under contract number MRTN-CT-2006-035606, the UK Science and Technology Facilities Council 
and by the Mexican Council of Science and Technology (CONACYT). Useful correspondence with Markus Schulze is acknowledged.}

\bibliographystyle{JHEP}

\end{document}